\documentclass[default,iicol]{sn-jnl}

\usepackage[T1]{fontenc}
\usepackage{graphicx}%
\usepackage{tabularx}
\usepackage{subcaption}
\usepackage{multirow}%
\usepackage{amsmath,amssymb,amsfonts}%
\usepackage{amsthm}%
\usepackage{mathrsfs}%
\usepackage[title]{appendix}%
\usepackage{xcolor}%
\usepackage{textcomp}%
\usepackage{manyfoot}%
\usepackage{booktabs}%
\usepackage{algorithm}%
\usepackage{algorithmicx}%
\usepackage{algpseudocode}%
\usepackage{listings}%

\newcommand\blfootnote[1]{%
\begingroup
\renewcommand\thefootnote{}\footnote{#1}%
\addtocounter{footnote}{-1}%
\endgroup
}

\theoremstyle{thmstyleone}%
\theoremstyle{thmstyletwo}%

\theoremstyle{thmstylethree}%

\begin{document}

\title[Article Title]{Characterisation of the neutron beam in the n\_TOF-EAR2 experimental area at CERN following the spallation target upgrade}


\author[1,2]{\fnm{J.A.} \sur{Pav\'{o}n-Rodr\'{\i}guez}} 
\author[3]{\fnm{J.} \sur{Lerendegui-Marco}}
\author[4,5]{\fnm{A.} \sur{Manna}}
\author[6]{\fnm{S.} \sur{Amaducci}}
\author[2*]{\fnm{M.} \sur{Sabat\'{e}-Gilarte}}
\author[7]{\fnm{E.} \sur{Musacchio-Gonzalez}}
\author[2]{\fnm{M.} \sur{Bacak}}
\author[8]{\fnm{V.} \sur{Alcayne}}
\author[1]{\fnm{M.A.} \sur{Cort\'{e}s-Giraldo}}
\author[2]{\fnm{V.} \sur{Vlachoudis}}
% ...

\affil[1]{Universidad de Sevilla, Spain}
\affil[2]{European Organization for Nuclear Research (CERN), Switzerland}
\affil[3]{Instituto de F\'{\i}sica Corpuscular, CSIC - Universidad de Valencia, Spain}
\affil[4]{Istituto Nazionale di Fisica Nucleare, Sezione di Bologna, Italy}
\affil[5]{Dipartimento di Fisica e Astronomia, Universit\`{a} di Bologna, Italy}
\affil[6]{INFN Laboratori Nazionali del Sud, Catania, Italy}
\affil[7]{INFN Laboratori Nazionali di Legnaro, Italy}
\affil[8]{Centro de Investigaciones Energ\'{e}ticas Medioambientales y Tecnol\'{o}gicas (CIEMAT), Spain}
% ...


\abstract{The n\_TOF facility at CERN has undergone a major upgrade after the installation of a new spallation target, designed to improve the performance of both neutron beamlines at the experimental areas 1 and 2 (EAR1 and EAR2) and the commissioning of a new experimental area (NEAR). Due to improved coupling of the spallation target with the EAR2 beamline, the upgrade resulted in a significantly increased neutron flux and improved neutron energy resolution. This paper presents the results of the commissioning phase that followed to characterise the EAR2 neutron beamline and validate the FLUKA Monte Carlo simulations of the facility. The main characteristics of the neutron beam, namely the neutron flux, spatial profile and energy resolution, are evaluated and compared to the previous target. The neutron flux presents a general increase of 20\% below  1~eV, 40\% between 1~eV and 100~keV and 50\% between 100~keV and 10~MeV. The measured width of the beam profile was 3~cm full width at half maximum (FWHM) at the reference position for neutron capture measurements. The energy resolution with the new spallation target shows a significant improvement compared to the previous one. Moreover, FLUKA Monte Carlo simulations present a good agreement with the measured neutron flux and profile within uncertainties, and a remarkable reproduction of the energy resolution.}

\keywords{n\_TOF Facility, Spallation neutron source, Neutron cross section, FLUKA}



\maketitle
\blfootnote{$^\ast$\textbf{Corresponding author:} \\ marta.sabate.gilarte@cern.ch \\ STFC – Rutherford Appleton Laboratory \\Particle Physics Department \\Harwell, Didcot OX11 0QX \\United Kingdom}
\section{Introduction}
\label{sec:intro} 

The n$\_$TOF Collaboration operates the neutron time-of-flight (TOF) facility at CERN, based on a 20~GeV/c pulsed proton beam impinging on a solid lead target where water is employed to moderate the neutrons produced by spallation reactions. The facility is characterized by a high-instantaneous neutron beam intensity, high energy resolution, and a wide neutron energy spectrum, spanning from sub-thermal to GeV. The scientific activities of the n\_TOF Collaboration are mostly focused on the measurement of neutron-induced cross sections of interest in astrophysics~\cite{Colonna2017}, nuclear technology~\cite{Colonna2020} and medical physics~\cite{PORRAS20142}. The first experimental area, EAR1, in operation since 2001, is located at 185 m from the spallation target nearly in the same direction as the incoming proton beam. In 2014, a second experimental area was commissioned, EAR2, located 20~m above the target perpendicular to the proton beam direction. 

During the CERN $2^{\textrm{nd}}$ Long Shutdown (2019-2021), the facility has gone through a major upgrade including the installation of a new spallation target designed to fully optimise the capabilities of the n$\_$TOF experimental areas, unlike the previous one specifically designed for EAR1~\cite{Guerrero2013}. Moreover, the development of new experimental area, NEAR, has been completed. NEAR \cite{Patronis2022} is located at 3~m to the left of the target with respect to the proton beam direction, aiming to explore neutron reactions of interest in astrophysics and radiation damage induced in materials by neutrons.  

The new target is expected to improve the performance of the EAR2 neutron beamline in terms of neutron flux and energy resolution. However, a good knowledge of every characteristic of the neutron beam is necessary for the correct analysis of the experimental data. For this purpose, an extensive commissioning of the neutron beam characterization has been carried out. 

In this paper, we present a brief description of EAR2 and the new spallation target (Section \ref{sec:2_ear2}) followed by a general description of the commissioning of the facility after its major upgrade. In Section \ref{sec:simulations} we introduce the key concepts of the Monte Carlo simulations of the facility, used for its design and later validated against measurements. Sections \ref{sec:flux}, \ref{sec:profile} and \ref{sec:RF} present the main characteristics of the neutron beam, i.e., the neutron flux, the beam profile and the energy resolution, respectively. Finally, Section \ref{sec:conclusions} gathers the summary and conclusions.

\section{EAR2: the n\_TOF 20~m neutron beamline}
\label{sec:2_ear2}

\begin{figure}[ht!]
    \includegraphics[width=0.5\textwidth]{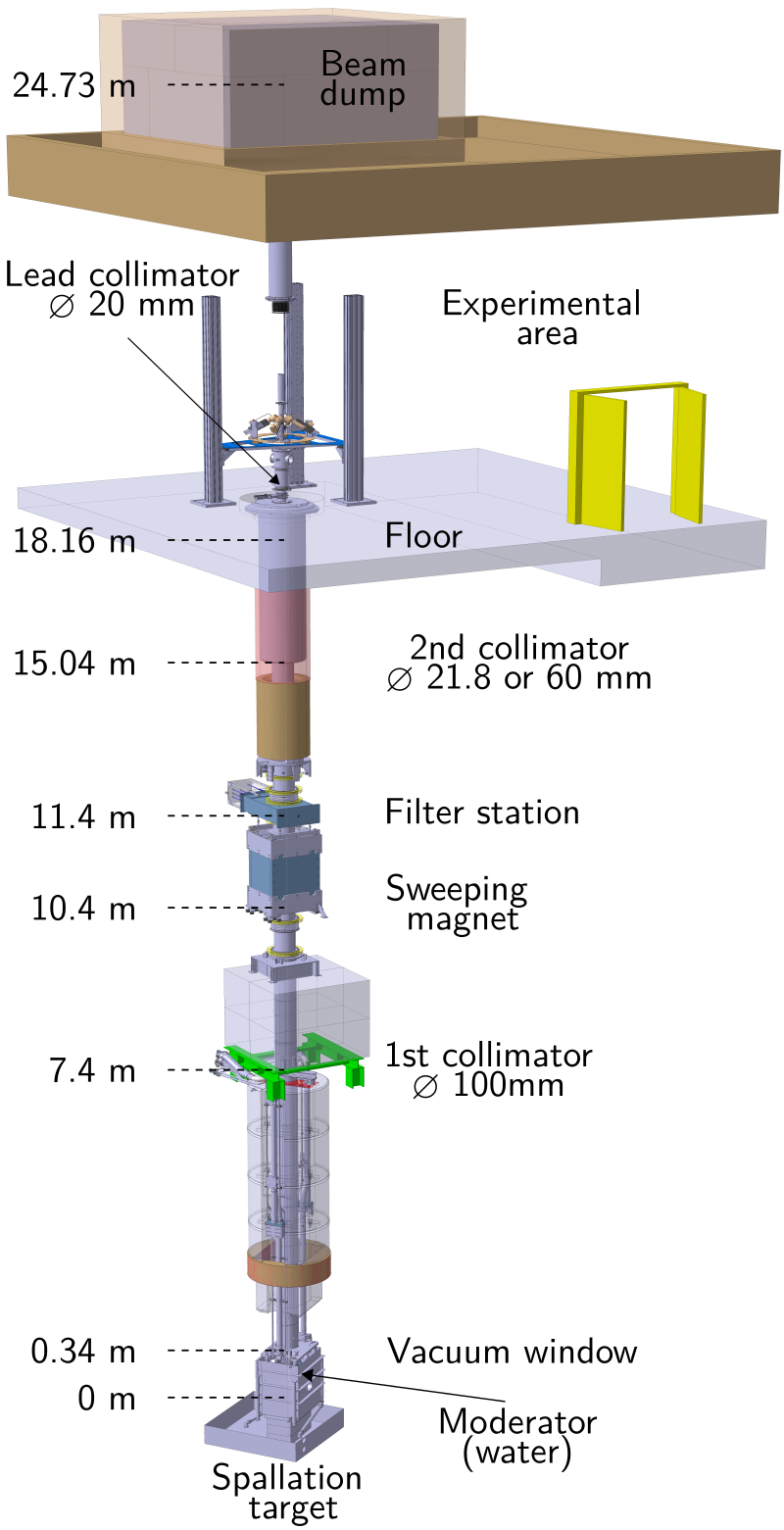}
    \caption{Vertical layout of the n\_TOF-EAR2 beamline (not to scale). Distances on the left indicate the upstream position of the elements with respect to the centre of the target.}
    \label{fig:1_beamline}       
\end{figure}

EAR2 is located above the spallation target at about 20~m along the vertical direction. Figure~\ref{fig:1_beamline} shows a drawing of the beamline's main components. From bottom to top: solid lead spallation target, water moderator, vacuum line window, first collimator, sweeping magnet, filter station, second collimator with two configurations depending on the experimental requirements (small and big, of 21.8 and 60~mm downstream inner diameter, respectively), the experimental area with a lead collimator (only in combination with the small collimator) and the beam dump. The second collimator has a conical shape, hence why the width of the beam profile widens after exiting the collimator (in more detail in Section \ref{sec:profile}). The first collimator is made of steel, while the second collimator is made of steel and polyethylene (including a boron carbide insert in the small configuration).

The EAR2 beamline was completed in 2014~\cite{Chiaveri2012,Weiss2015}, and it was in operation for 5~years using the existing spallation target that was optimised only for EAR1. During this phase, several measurements were performed~\cite{Barbagallo2016,Sabate2017conf,Stamatopoulos2020,Eleme2020,Alcayne2024Cm} exploiting the advantage of a high instantaneous flux, i.e., high number of neutrons per pulse, that allows measuring cross sections of samples of very low masses compared to EAR1, as the beam time required for the measurement is shorter and, most importantly, as the signal-to-background ratio for very radioactive samples is much higher. A comparison of pulsed neutron sources was carried out by Colonna et al. \cite{Colonna2018}, where EAR2 showed the highest neutron flux per proton pulse, with respect to the main facilities in operation worldwide. 

\begin{figure}
    \includegraphics[width=0.5\textwidth]{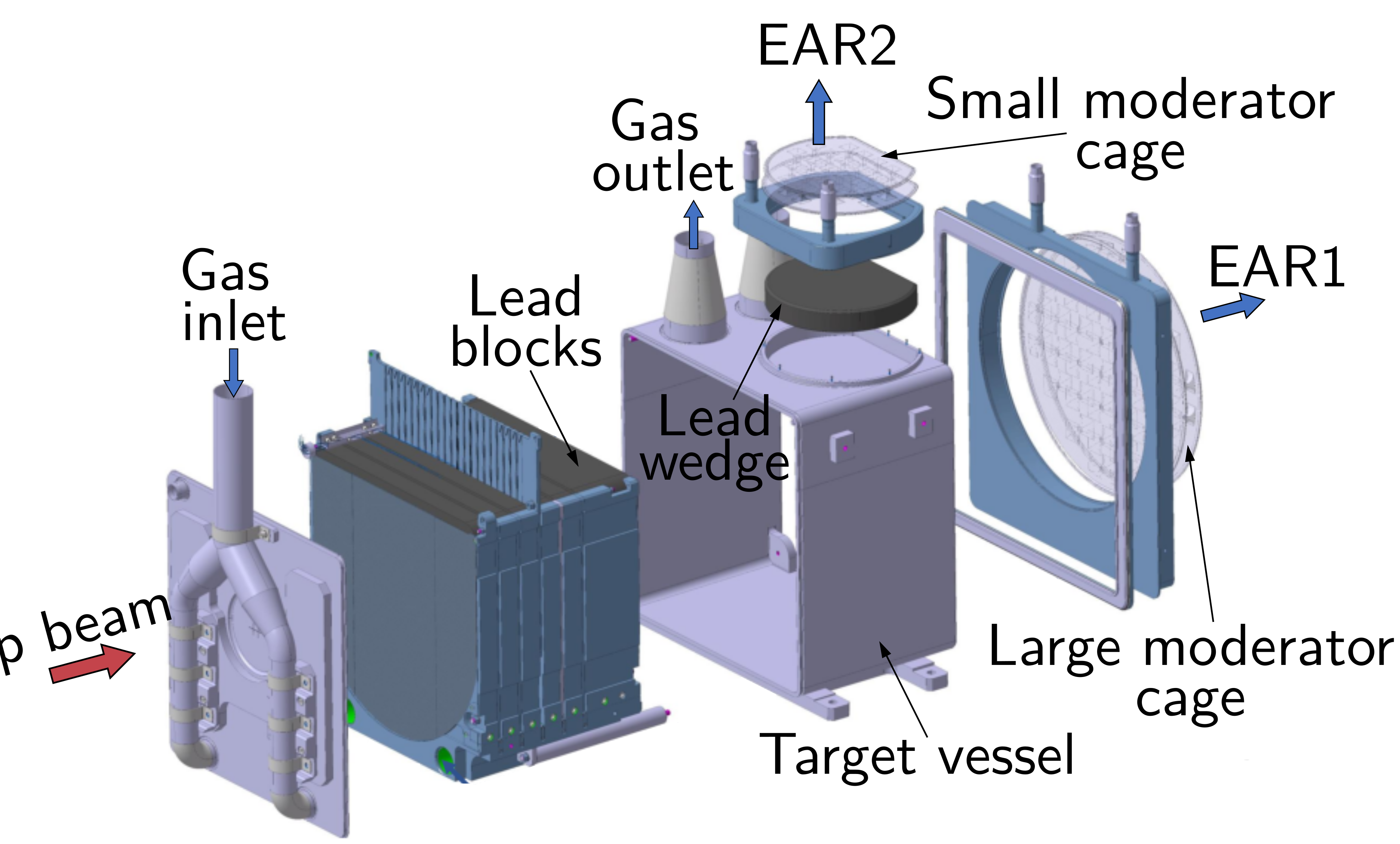}
    \caption{Exploded view (3D model) of the n\_TOF new spallation target~\cite{Esposito2020}. The red arrow indicates the direction of the impinging proton beam.
    }
    \label{fig:1target}      
\end{figure}

The new spallation target has been designed to provide a high-quality neutron beam at EAR2 without detriment to EAR1 beam. Figure~\ref{fig:1target} shows a 3D model exploded view of this target~\cite{Esposito2020}. The dedicated flat lead wedge and water moderator above such wedge (top right of the figure) are specifically designed to improve the energy resolution by means of a more isotropic production and moderation of the neutrons towards EAR2. In contrast, the previous spallation target was a monolithic lead cylinder coupled to EAR2 via a polygonal window~\cite{Weiss2015}.

\section{Monte Carlo simulations with FLUKA}
\label{sec:simulations}

Monte Carlo (MC) simulations allow to assess neutron beam characteristics such as the neutron flux and the spatial beam profile and are essential to determine the energy resolution of a TOF instrument over a wide range of neutron energies.
During the commissioning phase, the simulations are validated against experimental measurements, thereby becoming a powerful tool to further improve and optimise the quality of the neutron beam as well as other aspects of the experimental area such as the background conditions.
Moreover, they are indispensable in the accurate determination of the relation between energy and TOF, as discussed in detail in Sections \ref{sec:flux} and \ref{sec:RF}. 
In this section, we present the implementation and technical details of the MC simulations. 

\begin{figure}
    \includegraphics[width=0.5\textwidth]{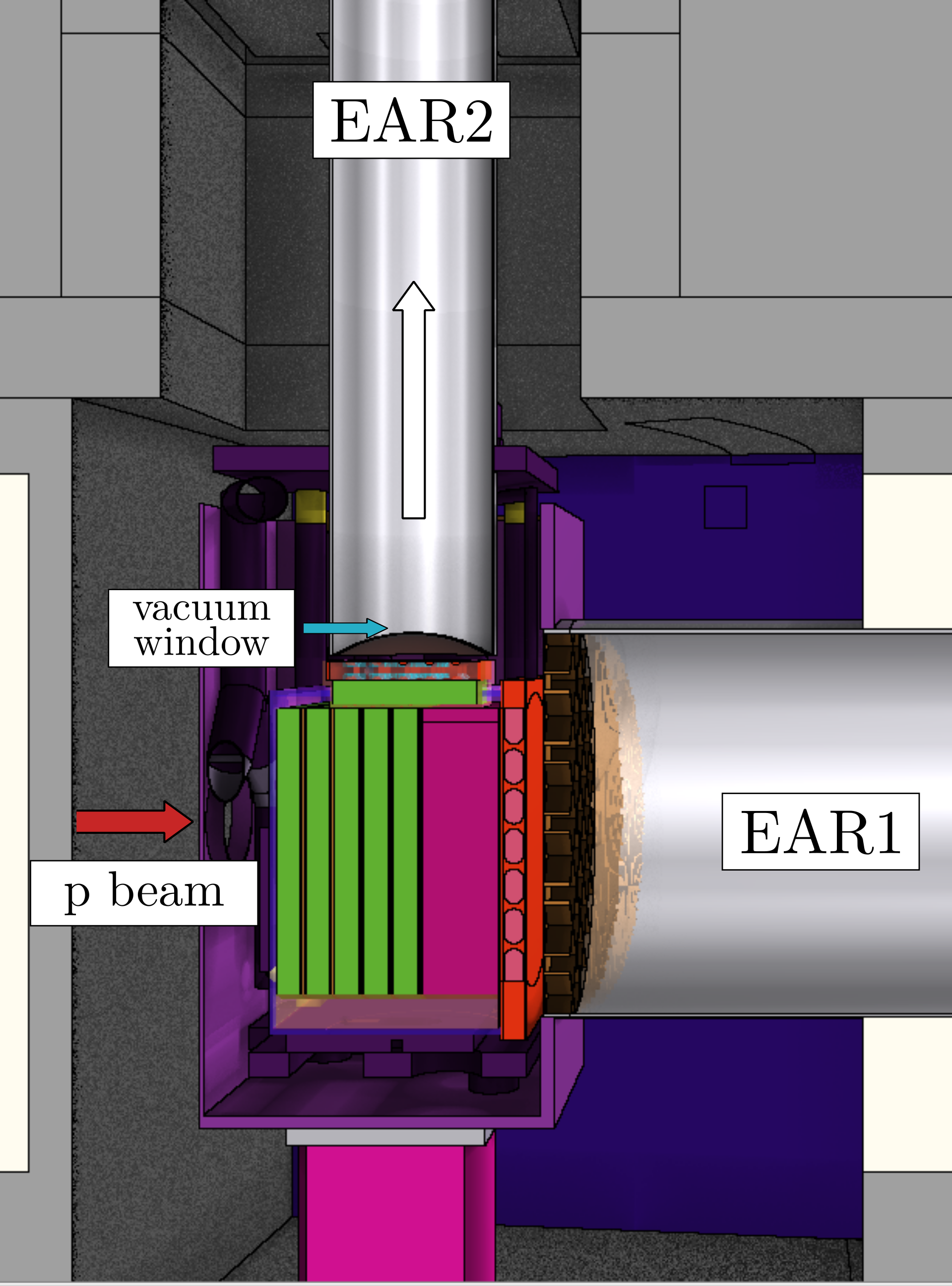}
    \caption{3D model of the FLUKA geometry of the target-moderator assembly generated with Flair, the advanced FLUKA graphical user interface.}
    \label{fig:2_FLUKA}       
\end{figure}

The simulation of the whole spallation process induced by protons impinging on the lead target is required to track and record the position, direction, energy and TOF of every neutron generated in the MC simulation and arriving in the experimental hall. The FLUKA code~\cite{webfluka,FLUKA1,FLUKA2} version 4.3 has been used to carry out these simulations. This version includes a point wise treatment of the neutron cross sections below 20~MeV~\cite{Vlachoudis2023}, while neutron interactions above this energy are managed by FLUKA's internal nuclear models, allowing a precise production and transport of the neutrons in the whole energy range of interest for n\_TOF. A detailed implementation of the geometry of the whole facility, based on its blueprints, has been carried out since the first phase of operation. For the new spallation target, the geometry of the previous phase has been adapted and updated accordingly, including a detailed description of all the constituent materials. In Figure~\ref{fig:2_FLUKA} we show a visualisation of the geometry implemented in FLUKA using its graphical user interface Flair~\cite{Vlachoudis2009}.

In terms of CPU time, these simulations are time-consuming and a full MC transport through the entire beamline is not feasible at the moment. To improve the speed of the simulations without compromising the reliability of the results, we have considered a two steps simulation approach. The positions, direction, energy and time relative to the generation of the neutrons and $\gamma$-rays emitted from the target in the spallation process are scored on the hemispherical surface of the vacuum pipe window, at the interface from the aluminium window to vacuum, that is indicated with a blue arrow in Figure~\ref{fig:2_FLUKA}. Within a sufficiently small angle ($\theta_{\rm cut} = 5^{\circ}$ for EAR2~\cite{Bergstrom2013}), relative to the neutron beam pipe axis, the neutron's angular distribution can be assumed to be isotropic. We continue the neutron transport up to the experimental hall by optically propagating the neutrons along the beamline, performing a sampling of the incident angle within $\theta_{\rm cut}$. This optical transport is carried out by a C++ custom code~\cite{Bergstrom2013, Vlachoudis2021}. The collimation system is modelled in such a way that, if a neutron hits the collimator, it is discarded and does not arrive in the experimental hall. In this way, we can also reproduce the spatial distribution of the neutron beam as shown in Section \ref{sec:profile}. The neutrons arriving to the experimental area at the requested flight path are computed according to their position, energy and TOF to obtain the quantities of interest.  

\section{Neutron flux}
\label{sec:flux}

The definition of flux, $\Phi (E_\textrm{n})$, according to the International Commission of Radiation Units and Measurements (ICRU)~\cite{ICRU85a} is
\begin{equation}
	\Phi_{\rm ICRU} (E_\textrm{n}) = \frac{dN}{dtdE_\textrm{n}},
\end{equation} 
where $dN$ is the increment of the number of neutrons with energy between $E_\textrm{n}$ and $E_\textrm{n}+dE_\textrm{n}$ in the time interval $dt$, with units of $\textrm{s}^{-1} \textrm{J}^{-1}$.

At n\_TOF the measurements are normalised per total number of protons on target rather than per duration of the measurement in seconds, since the proton pulse frequency is not constant and can differ considerably from the average 0.8~Hz. Therefore, the definition of flux used at n\_TOF is \textit{the number of neutrons with energy between $E_\textrm{n}$ and $E_\textrm{n}+dE_\textrm{n}$, integrated over the entire profile of the beam arriving to the experimental area and per proton on target}, 
\begin{equation}
	\Phi (E_\textrm{n}) = \frac{dN}{dE_\textrm{n}},
\end{equation}
with units of $\textrm{J}^{-1}$ and scaled to the nominal proton pulse, which is produced by $7\times 10^{12}$ protons.

A precise knowledge of the neutron flux at the sample position is required for achieving highly accurate measurements. For this reason, we have carried out a commissioning campaign measurement of the flux combining several complementary detection systems and different standard cross sections. We refer to \textit{flux evaluation} to this way of determining the flux, and \textit{evaluated flux} to the high precision measured flux, in order to differentiate from the often less precise flux monitored online during the measurements.  

\subsection{Methodology}

\begin{figure}
    \centering
    \includegraphics[width=1\linewidth]{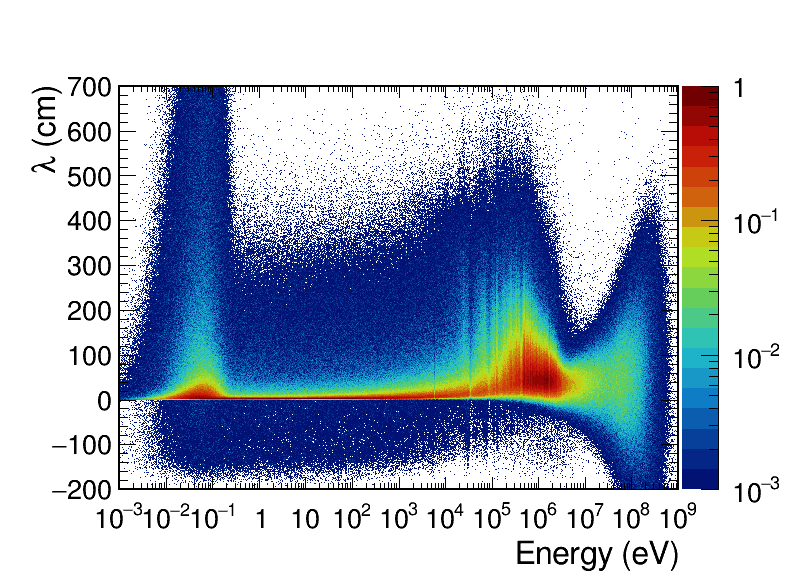}
    \caption{FLUKA simulated $\lambda$ distribution as a function of $E_\textrm{n}$.}
    \label{fig:4_T2E_RF}
\end{figure}

The reaction yield is defined as the fraction of neutrons that cause a specific nuclear reaction occurring as a function of the incident neutron energy $E_\textrm{n}$. The theoretical yield, $Y^{\rm th}(E_{\rm n})$, is defined as
\begin{equation}
    Y^{\textrm{th}}(E_\textrm{n})=(1-e^{-n\sigma_\textrm{t}(E_\textrm{n})})\frac{\sigma_\textrm{r}(E_\textrm{n})}{\sigma_\textrm{t}(E_\textrm{n})},
    \label{eq:thyield}
\end{equation}
where $n$ is the areal density of the sample, $\sigma_r (E_{\rm n})$ and $\sigma_t (E_{\rm n})$ are the reaction and total cross section, respectively. 
Experimentally, it can be determined as
\begin{equation}
    Y^{\textrm{exp}}(E_\textrm{n})=\frac{C(E_\textrm{n})-B(E_\textrm{n})}{\varepsilon(E_\textrm{n})\cdot\Phi(E_\textrm{n})},
    \label{eq:expyield}
\end{equation}
where $C(E_\textrm{n}$) is the number of counts registered, $B(E_\textrm{n}$) is the background contribution, and $\varepsilon(E_\textrm{n})$ is the total detection efficiency, all of them expressed per energy unit and integrated per nominal proton pulse. Combining equations (\ref{eq:thyield}) and (\ref{eq:expyield}), it is possible to extract the neutron flux: 
\begin{equation}
    \Phi(E_\textrm{n})=\frac{C(E_\textrm{n})-B(E_\textrm{n})}{\varepsilon(E_\textrm{n})\cdot(1-e^{-n\sigma_\textrm{t}(E_\textrm{n})})\frac{\sigma_\textrm{r}(E_\textrm{n})}{\sigma_\textrm{t}(E_\textrm{n})}}.
    \label{eq:flux}
\end{equation}
At n\_TOF, for each isotope, $C(E_\textrm{n})-B(E_\textrm{n})$ is not measured as a function of $E_\textrm{n}$ but as a function of the TOF, $T$. $T$ is defined as the time interval between the time of signal detection, $t$, and the time when neutrons escape the spallation target. In practice, the latter is determined from the arrival time of the $\gamma$ rays, $t_{\gamma}$, generated in the spallation process (also known as the $\gamma$-flash~\cite{Weiss2015}), corrected for the time it takes these $\gamma$ rays to travel to the sample position, i.e., 
\begin{equation}
T= t - t_{\gamma} + \frac{L_0}{c},
\label{eq:measure_tof}    
\end{equation}
where $c$ is the speed of light and $L_0$ is the flight path (distance from the moderator-target assembly to the experimental hall). After applying the necessary efficiency corrections to each detection system, we obtain the experimental yield as a function of $T$. $E_\textrm{n}$ is related with $T$ via
\begin{equation}
E_\textrm{n}= m_\textrm{n} \cdot c^2 \cdot \left(\frac{1}{\sqrt{(1-(\frac{L_0+\lambda}{c \cdot T})^2)}}-1\right),
    \label{eq:tof2e}
\end{equation} 
where $m_\textrm{n}$ is the neutron rest mass and $\lambda$ is an effective neutron path before reaching the hemispherical vacuum window, that accounts for the stochastic nature of the moderation process \cite{Vlachoudis2021}. In fact, $\lambda$ is a stochastic quantity with a distribution shown in Fig \ref{fig:4_T2E_RF}. It means that a distribution of $E_\textrm{n}$ corresponds to a specific measured $T$, or conversely, neutrons with the same $E_\textrm{n}$ arrive at a different $T$. The relation of $E_\textrm{n}$ and $T$ (or $\lambda$) is then called energy {\em resolution function} (RF) of the facility and is discussed in detail in Sec.~\ref{sec:RF}.

According to Eq. (\ref{eq:flux}), the neutron flux is extracted by dividing the number of counts (per proton pulse) by the theoretical yield. In the latter, any involved cross section is convoluted with the $\lambda$ distribution, obtained via MC simulations of the spallation process carried out with FLUKA, in order to account for the effect of the RF.
The flight path in Eq. (\ref{eq:tof2e}) is determined by means of an iterative process where the position of the resonances in the $Y^\textrm{exp}$ in time is compared to the ones in the theoretical yield convoluted with the RF. In the absence of resonances, the dips caused by the neutron absorption in the aluminium windows of the target can serve for the same purpose. 
The result of this process\textemdash when the resonances (or dips) match in both yields\textemdash gives an effective flight path, that accounts for the effect of the RF. Typically, the same flight path can be used for the whole measured energy. However, there are cases in which multiple flight paths need to be used for different energy ranges. 

\subsection{Flux evaluation}
The experimental setup consisted of solid state detectors, SiMon2 (Silicon Monitor)~\cite{SiMon2}, and gaseous detectors, MGAS (Micro-mesh Gaseous Structure)~\cite{MGAS1,MGAS2,MGAS3} and PPACMon (Parallel Plate Avalanche Counters)~\cite{PPAC1,PPAC2}. The details on the analysis of the different reactions is specified in the detector references. The use of different detector systems aims to reduce the systematic uncertainties. The commissioning measurements were carried out with the small collimator, since it is the most frequently used one. 
\begin{figure}
    \includegraphics[width=0.5\textwidth]{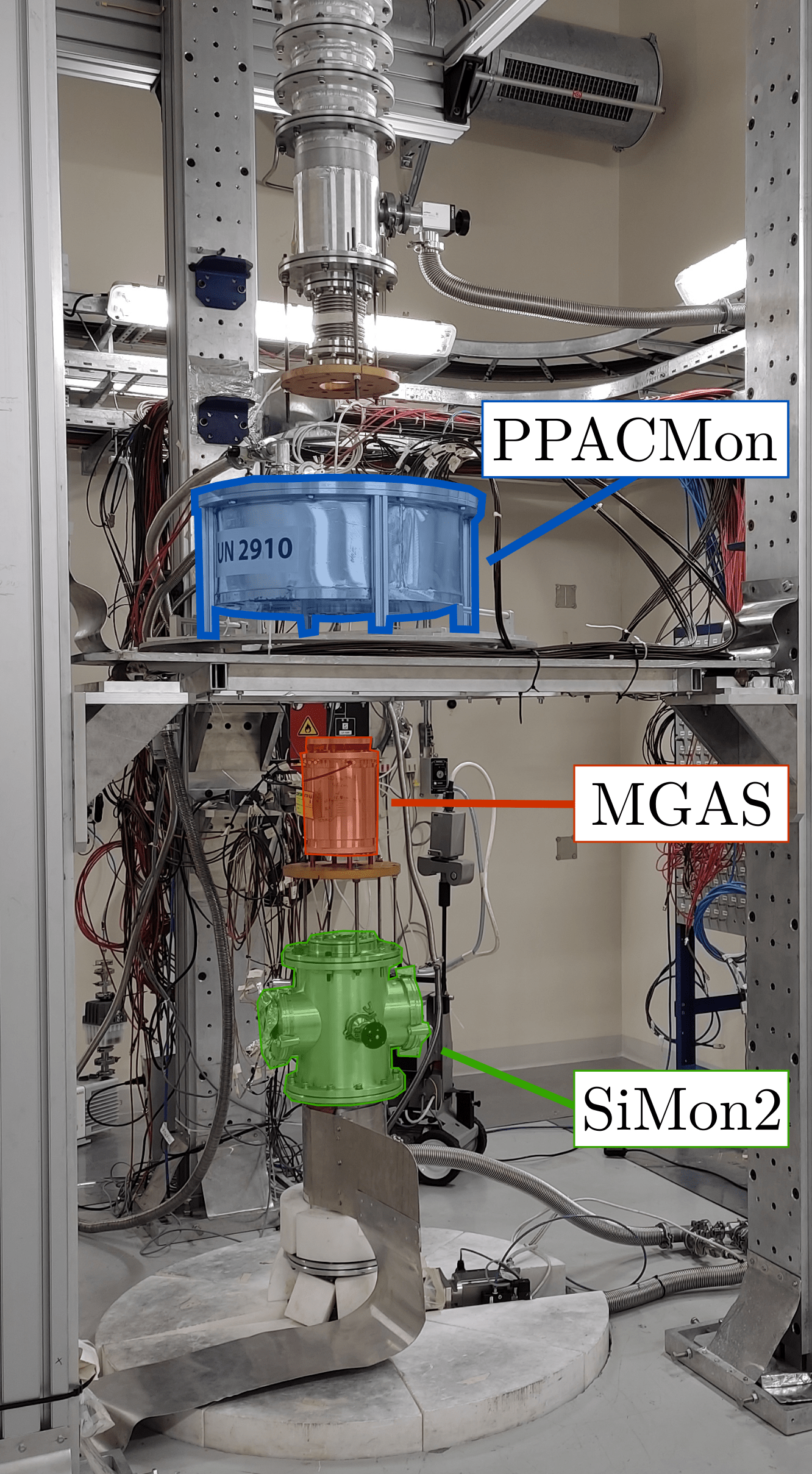}
    \caption{Experimental setup for the determination of the neutron flux at EAR2, consisting of SiMon2 (green), MGAS (red) and PPACMon (blue), placed in upstream order.}
    \label{fig:4setupFlux}       
\end{figure}
Figure~\ref{fig:4setupFlux} shows the setup in the experimental hall. The neutron beam comes from the bottom, passing through SiMon2, MGAS and PPACMon before being stopped at the beam dump downstream (not shown), placed above the ceiling. The isotopes, reactions and sample masses used for every detector, together with the ranges of energy in which they have been used are summarised in Table~\ref{tab-ranges}. The sizes of all the samples employed in the flux evaluation are large enough to cover the whole neutron beam profile. The energy region from 25~meV to 200~keV is covered by SiMon2 with $^6$Li and from 25~meV to 30~keV by MGAS with $^{10}$B. The $^{6}$Li(n,t) and $^{10}$B(n,$\alpha$) capture cross sections are considered standard between 25~meV and 1~MeV~\cite{Carlson2018}, although due to the effects of the $\gamma$-flash and pile-up in our measurements, the first is limited to 200~keV and the latter to 30~keV. In particular, in the case of $^{6}$Li(n,t), the effect of the angular distribution of the reaction above 100~keV was taken into account in the estimation of the detector efficiency via simulations. The values of the flux below 25~meV presented are also obtained by a combination of these two datasets, since there were no other standards available. The $^{235}$U(n,f) cross section is considered standard within the energy range of 0.15 - 200~MeV; however, we have decided to extend the use of $^{235}$U data down to 30~keV, since the $^{235}$U(n,f) cross section is smooth and well-known in this energy range. MGAS~$^{235}$U is reliable up to ~5~MeV because of the $\gamma$-flash, and from 1~MeV it uses PPACMon $^{235}$U data as a TOF reference due to the impact of the slower MGAS signals on the initial time determination in such energies. The energy range from 5~MeV up to 200~MeV is then covered only with PPACMon $^{235}$U and $^{238}$U, while above 200~MeV the count rate is very low for achieving reliable statistics. Furthermore, since the $^{235}$U sample of PPACMon has the best-characterised mass, with an accuracy of only 0.4\%, it has been used for absolute normalization of the flux. In particular, the flux measured with all the detectors has been normalised to the integral of the PPACMon flux in the 7.8-11~eV range, which for $^{235}$U(n,f) is also considered a standard~\cite{Carlson2018}.
\begin{table} [t]
\centering
\caption{List of isotopes, reactions and sample masses for each detector and their corresponding energy ranges of interest.}
\label{tab-ranges}       
\begin{tabular}{cccc}
\hline
Detector    & Reaction                 & Mass                 & $E_{\textrm{n}}$ range \\
            &                        & ($\mu$g/cm$^2$)       & \\
\hline
 SiMon2      & $^{6}$Li(n,t)          & 78.8                 & 25 meV - 200 keV \\ 
 MGAS       & $^{10}$B(n,$\alpha$)   & 6.1                  & 25 meV - 30 keV \\
            & $^{235}$U(n,f)         & 117.6                & 30 keV - 5 MeV \\ 
 PPACMon    & $^{235}$U(n,f)         & 280.0                & 30 keV - 200 MeV \\
            & $^{238}$U(n,f)         & 17                    & 3 - 200 MeV      \\
\hline
\end{tabular} 
\end{table}
The evaluated neutron flux is obtained from the combination of the different measurements in the energy regions described. For this, we have done a weighted average of the measurements, where the weights have been defined according to the statistical uncertainty. To verify the absolute value of the neutron flux we have also carried out a measurement of the flux at 4.9~eV from the activation of $^{197}$Au foils of 100~$\mu$m thickness using the saturated resonance technique~\cite{Macklin1979}.  
\begin{figure*}[t]
\centering
\includegraphics[width=1.\textwidth]{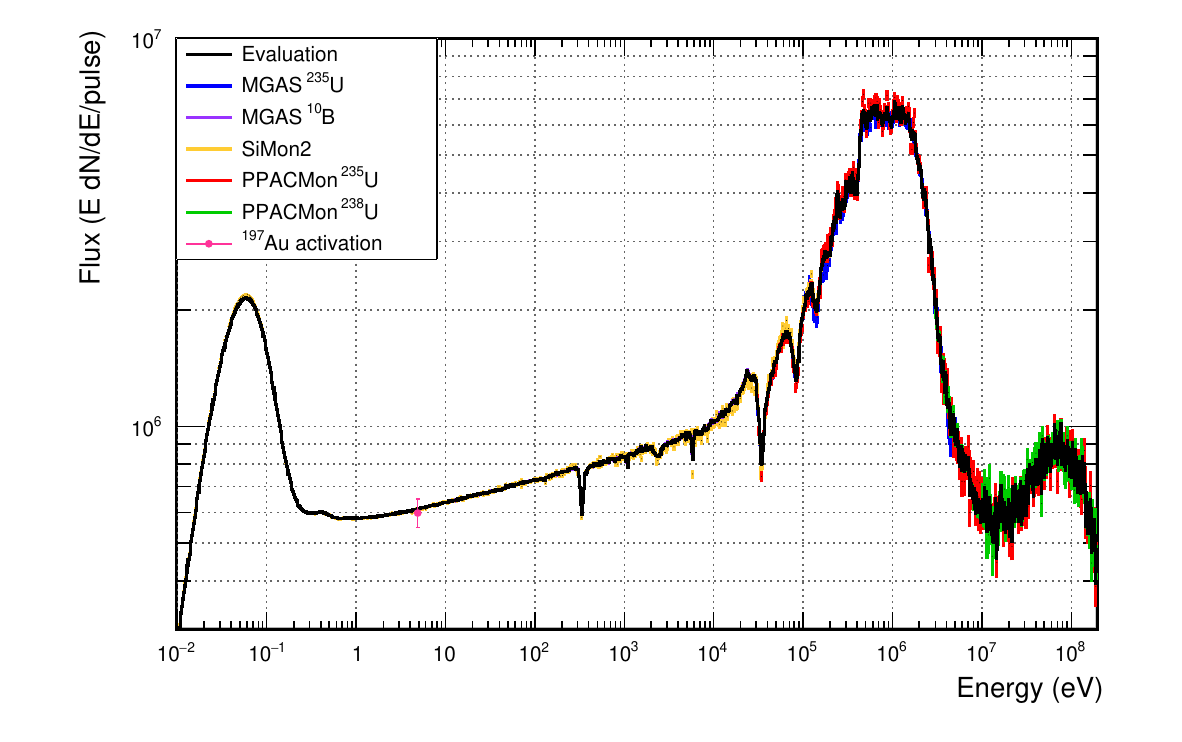}
\caption{EAR2 evaluated neutron flux (black line) together with different detector measurements (colour lines) in their corresponding energy ranges of use and the activation measurement (pink dot). The histograms are represented in isolethargic units and 100 bins per decade. One pulse corresponds to $7 \times10^{12}$ protons.}
\label{fig:4fluxEva}    
\end{figure*}

Figure~\ref{fig:4fluxEva} shows the evaluated flux (in units of lethargy) together with the results from different detectors, in the energy regions in which they have been used, and the activation measurement. The latter is in perfect agreement with the evaluation. 
\begin{figure}
    \includegraphics[width=0.5\textwidth]{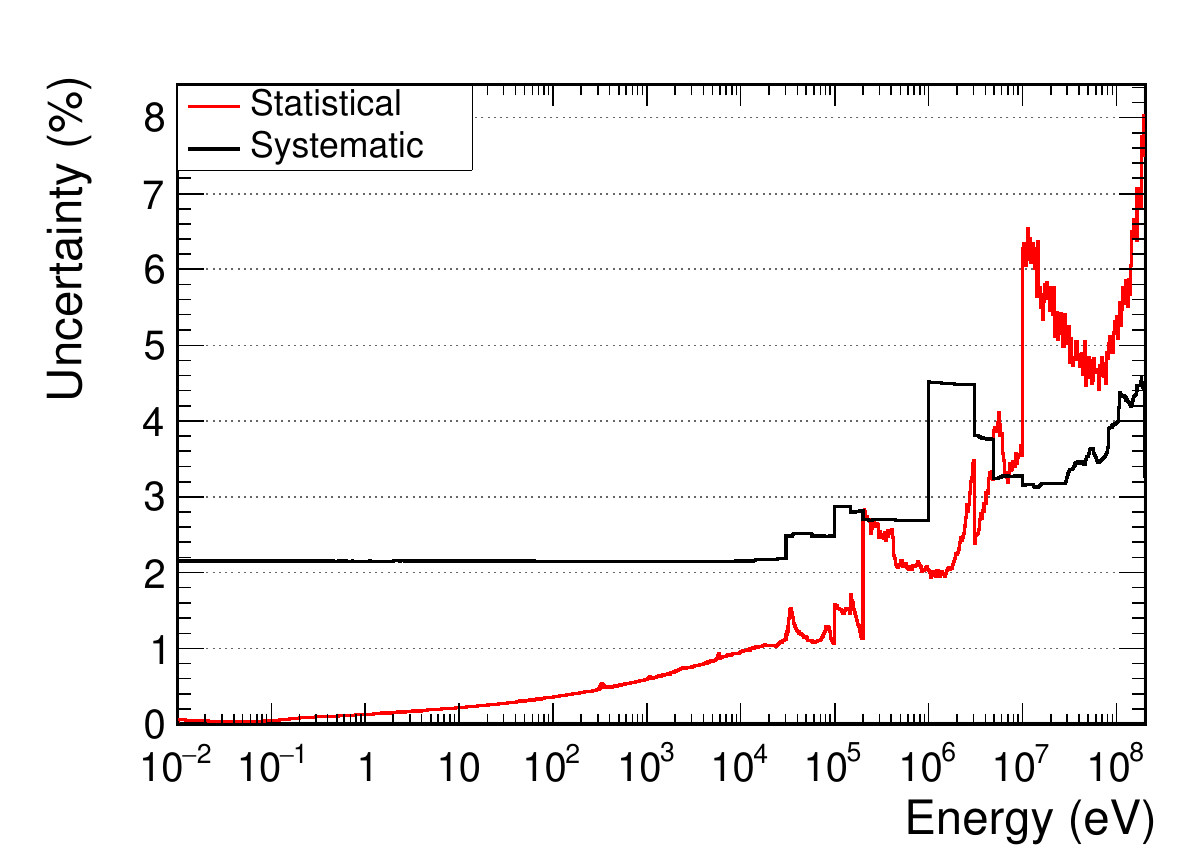}
    \caption{ Statistical (red) and systematic (black) uncertainties of the evaluated neutron flux in 100 bins per decade.}
    \label{fig:4fluxunc}       
\end{figure}
Figure~\ref{fig:4fluxunc} shows the statistical and systematic uncertainties as a function of the neutron energy $E_{\rm n}$. The systematic uncertainty for each of the measurements has been calculated by means of error propagation, considering the uncertainty of the different corrections applied during the data analysis, as well as the uncertainties related to the use of evaluated cross sections and MC simulations. Above 10~MeV the total number of neutrons arriving into the experimental hall decreases considerably. Therefore, the difficulty of detecting neutrons is bigger and the statistical error dominates the uncertainty. The jump in statistical uncertainties at 10~MeV is due to the use of a single detector in the evaluation in this range. The same effect can be seen in the systematic uncertainties. As we go down in energy, the systematic uncertainties are more relevant. Below 10~keV the systematic uncertainties come mainly from the MC simulations used in the calculation of the flux. This uncertainty is the same in the whole energy range. The flux derived from different detectors is also fully consistent within the statistical uncertainties.
The evaluated flux presents a series of dips between 50 and 200~keV, due to neutron transmission through the aluminium windows in the target-moderator assembly and the beam line, which present resonances in its cross sections. The effect of these dips is minimal for the measurement of actinide cross sections in their unresolved resonance region, while for measurements of cross sections with resonances this energy range is already constrained by the RF. 
Table~\ref{tab:fluxvalues} shows the integral value of the neutron flux per proton pulse in each decade of the energy spectrum and the average of the systematic and statistical uncertainties in the same decade. 
\begin{table} [t]
\centering
\caption{Neutron flux value for a nominal intensity of 7$\times 10^{12}$ protons per pulse integrated over each neutron energy decade. The statistical and systematic uncertainties are also presented.}
\label{tab:fluxvalues}
\begin{tabular}{lcrr}
\hline
Neutron & Neutrons      & Stat. (\%) & Syst. (\%) \\
energy         & per pulse           &         &   \\\hline
 10-100 meV    & $3.14 \times 10^6 $  & $<$ 0.1  &  2.2 \\ 
 0.1-1  eV     & $1.69 \times 10^6 $  & 0.1 & 2.2   \\ 
 1-10   eV     & $1.40 \times 10^6 $  & 0.2 & 2.2   \\ 
 10-100 eV     & $1.58 \times 10^6 $  & 0.3 & 2.2   \\ 
 0.1-1 keV     & $1.80 \times 10^6 $  & 0.5 & 2.2   \\ 
 1-10 keV      & $2.11 \times 10^6 $  & 0.8 &  2.2  \\ 
 10-100 keV    & $3.07 \times 10^6 $  & 1.1 &  2.4  \\ 
 0.1-1 MeV     & $1.00 \times 10^7 $  & 2.2 & 2.8   \\ 
 1-10 MeV      & $6.77 \times 10^6 $  & 2.9 & 4.0   \\ 
 10-100 MeV    & $1.66 \times 10^6 $  & 5.3 & 3.4   \\
100-200 MeV    & $4.61 \times 10^5 $  & 6.5 & 4.4  \\\hline
\end{tabular}
\end{table}

\begin{figure}
    \includegraphics[width=0.5\textwidth]{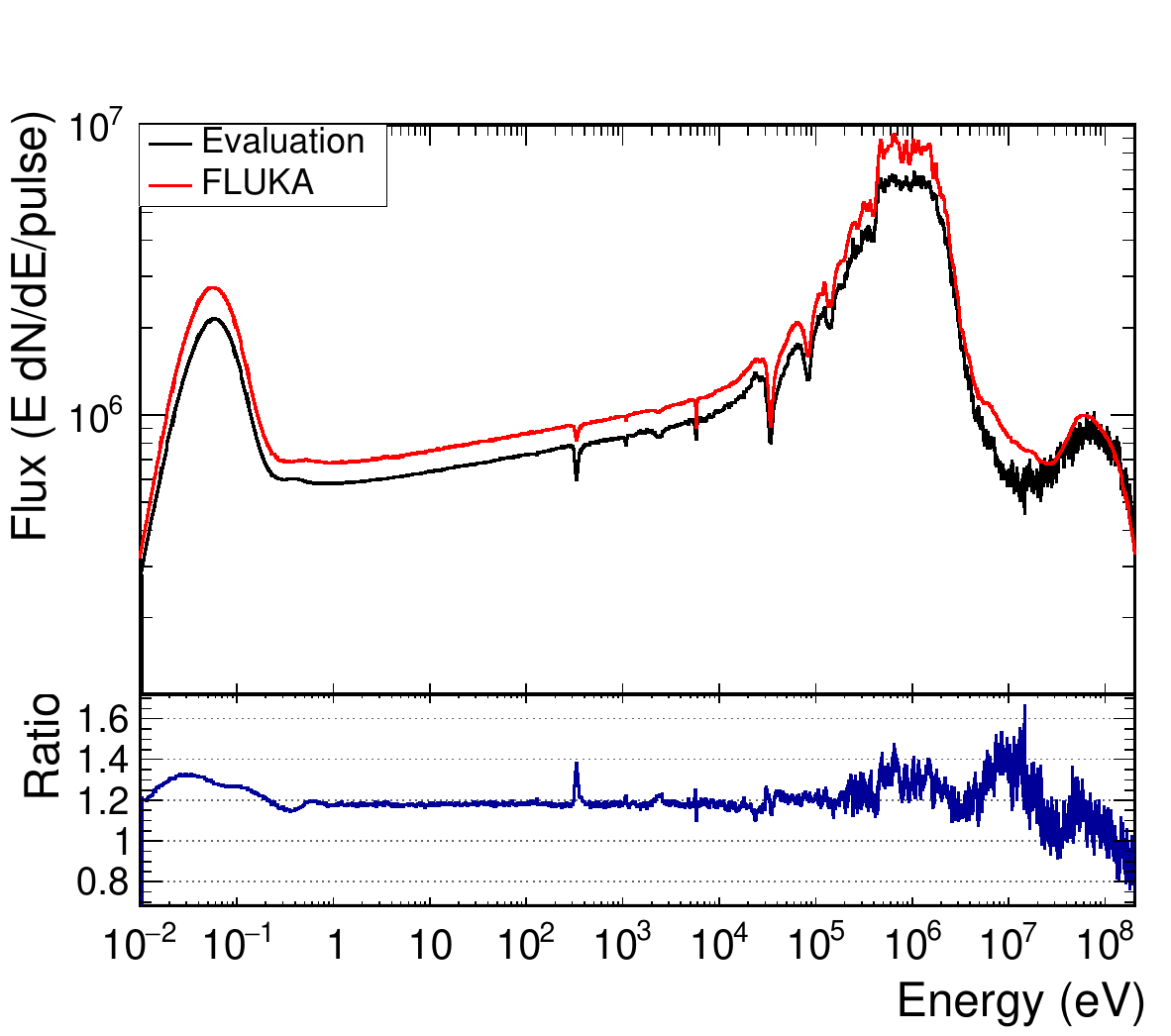}
    \caption{Top panel: Evaluated neutron flux (black) and FLUKA simulated flux (red). Bottom panel: ratio between FLUKA and the evaluated flux.}
    \label{fig:4fluxfluka}     
\end{figure}

Figure~\ref{fig:4fluxfluka}, top panel shows the evaluated neutron flux along with the results of the FLUKA simulations carried out in two steps (production of spallation neutrons, and resampling and propagation up to the experimental hall). The bottom panel then shows the ratio between FLUKA and the evaluated flux. The absolute value of the simulated flux is about 20\% higher with respect to the experimental data, similar ratio as obtained with the previous spallation target ~\cite{Sabate2017}. The shape of the neutron energy flux is in general well reproduced by the simulations, which after scaling to the evaluation, are on average in agreement within a 3\% per energy decade in the range from 1~eV to 100~keV, while in the regions below 1~eV and from 100~keV up to 20~MeV the average agreement is within a 10\%. These differences might arise from possible imperfections in the modeling of the geometry and materials, as well as small misalignments in elements in the actual beamline such as collimators. Above 20~MeV the simulations underestimate the experimental flux, an effect that we also observe in EAR1 and NEAR. In this energy range the simulations rely only on theoretical models of the neutron generation and transport, and the origin of the inconsistencies is unclear. 

\subsection{Comparison with the previous spallation target}

\begin{figure}
    \includegraphics[width=0.5\textwidth]{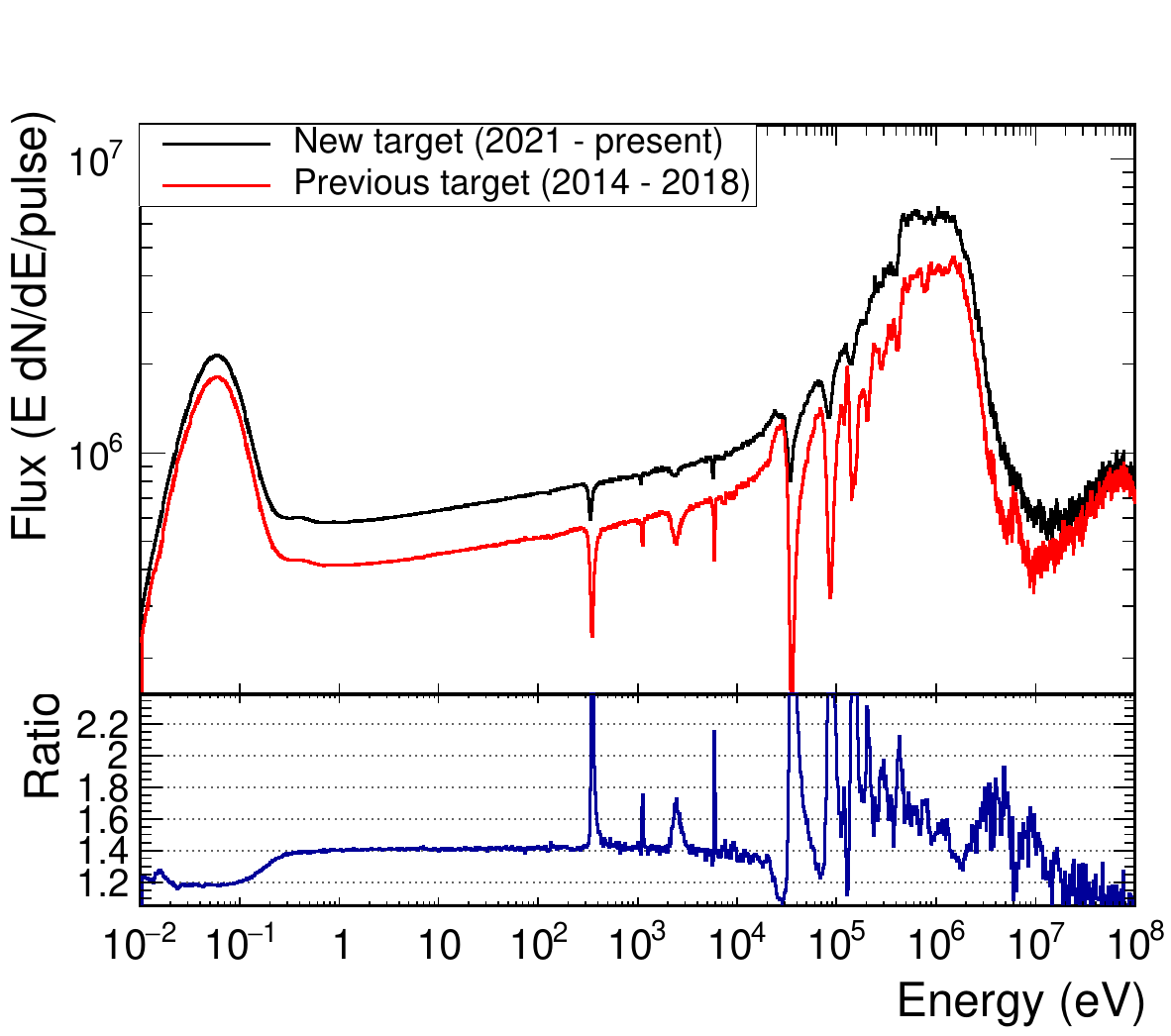}
    \caption{Top panel: Evaluated flux with the new (black) and previous (red) target targets. Bottom panel: ratio of evaluated flux with the new and previous targets.}
    \label{fig:4_p3n4}       
\end{figure}

As a result of the redesign and installation of the new spallation target, a general increase in the absolute value of the neutron flux can be observed in the whole energy range. The results of the neutron flux evaluation for the previous target (2014-2018)~\cite{Sabate2017}, are compared to the one with the new target (2021-present) in Figure~\ref{fig:4_p3n4}. In particular, an increase at average of 40\% is observed between 1~eV and 100~keV, while of about 20\% below 0.5~eV. From 100~keV to 10~MeV the increase is 50\%. This further increases the performance of measurements with small mass samples and/or small cross sections, and improve the signal to background ratio for very radioactive samples at EAR2, with respect to the previous spallation target. A comparison with the neutron flux in EAR1 using the new target is planned for a forthcoming work, currently under preparation, focused on the MC simulations of both areas.

\section{Spatial beam profile}
\label{sec:profile}

\begin{figure}
    \includegraphics[width=0.5\textwidth]{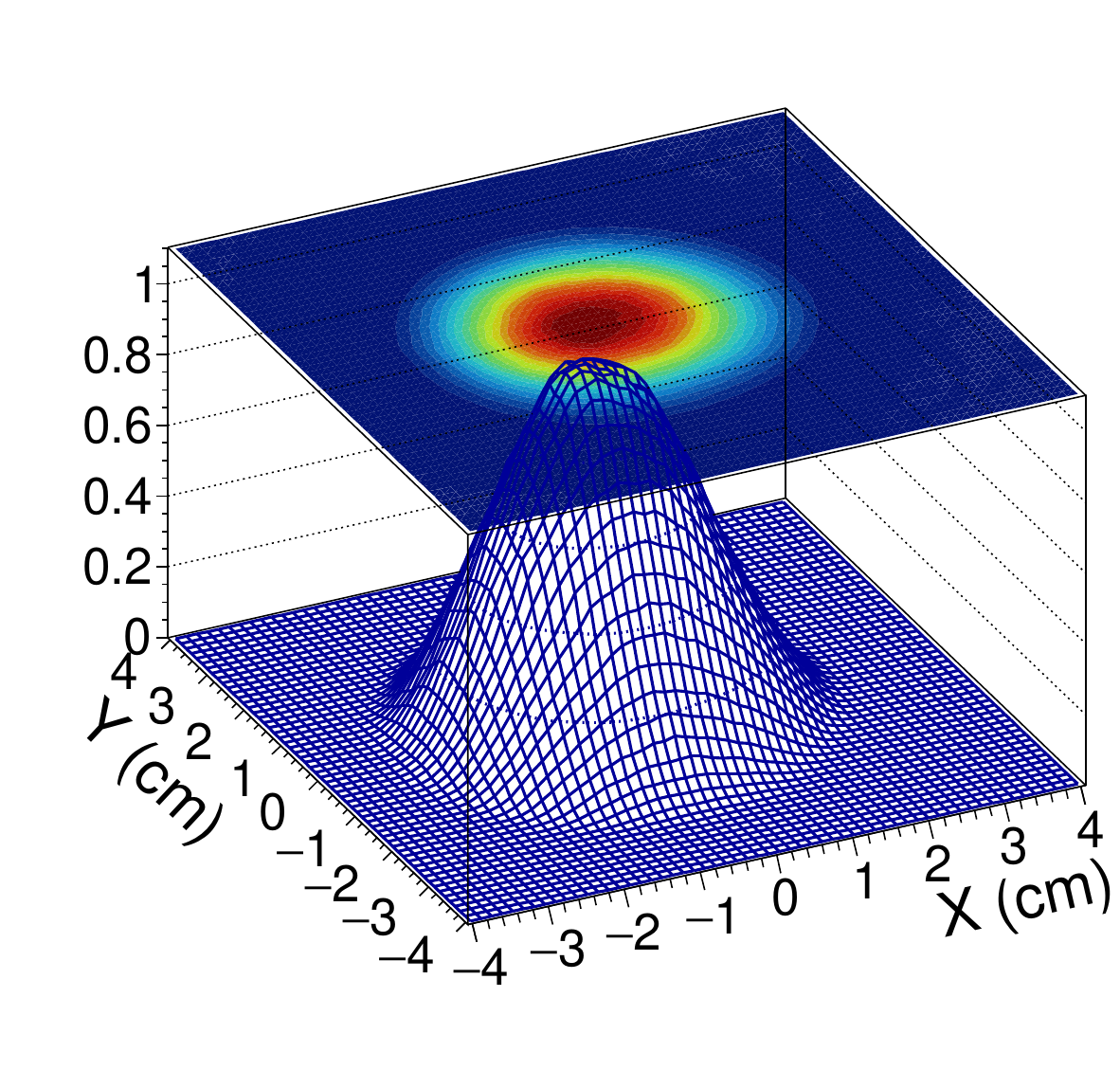}
    \caption{Neutron beam profile integrated over all neutron energies measured with PPACMon at 19.95~m flight path. The maximum of the distribution is scaled to 1.}
    \label{fig:5_ppac_profile}       
\end{figure}

The beam profile with the small collimator was determined from PPACMon measurements profiting from the 1.5~mm spatial resolution of this detector. Figure~\ref{fig:5_ppac_profile} shows the neutron beam profile measured at a flight path $L=19.95$~m and integrated over the whole energy range. The Y-axis corresponds to the direction of the proton beam, i.e., perpendicular to the spallation target, while the X-axis is parallel to the target, the Z-axis being oriented along the EAR2 beamline. In practice, the proton beam impinges on the target at a 10$^\circ$ angle relative to the Y-axis, directed toward the negative X-axis, to reduce EAR1's background due to $\gamma$ rays and charged particles. 

\begin{figure}
\centering
\begin{subfigure}{.5\textwidth}
\centering
    \includegraphics[width=.9\linewidth]{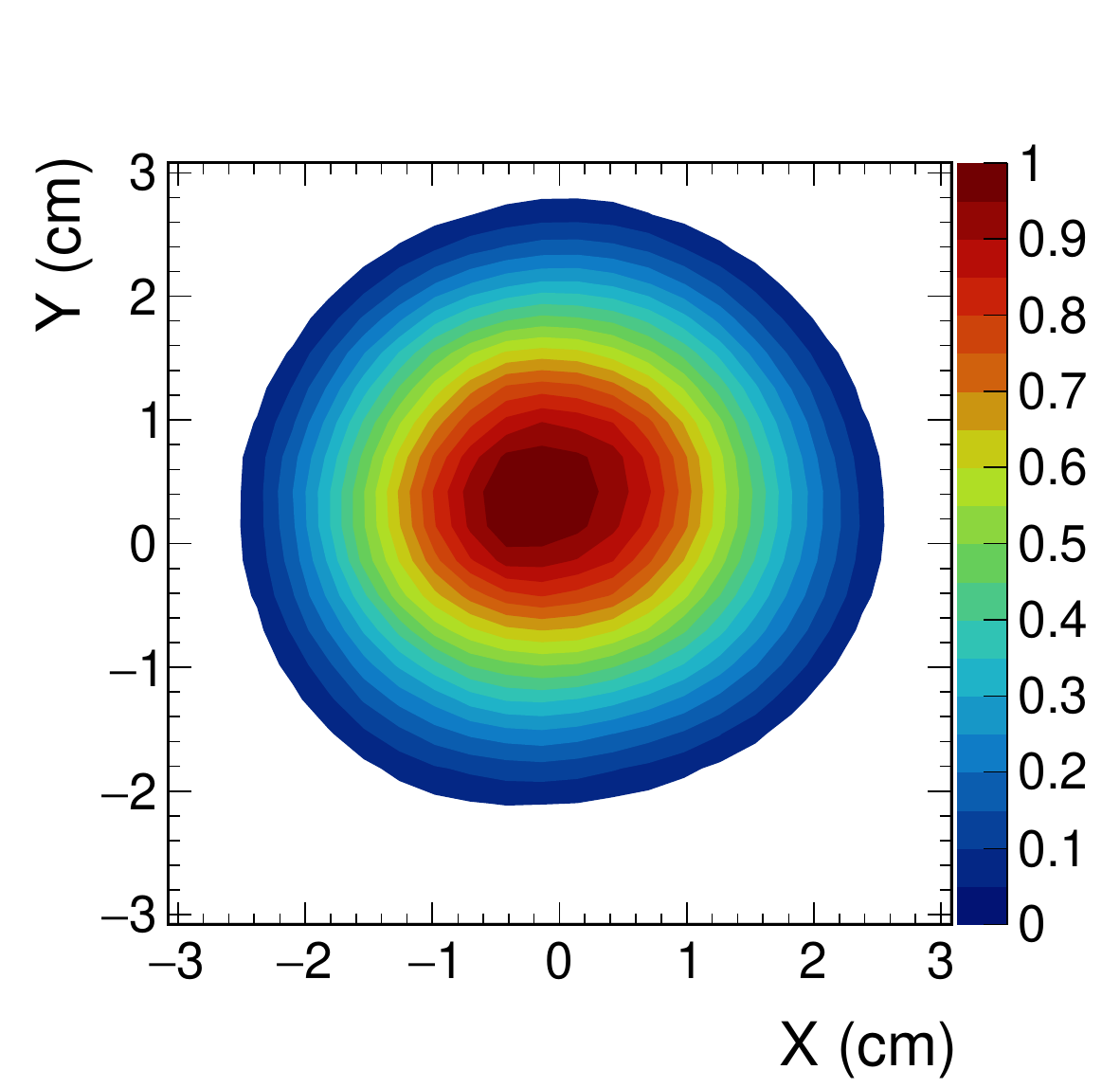}
\end{subfigure}
\begin{subfigure}{.5\textwidth}
\centering
    \includegraphics[width=.9\linewidth]{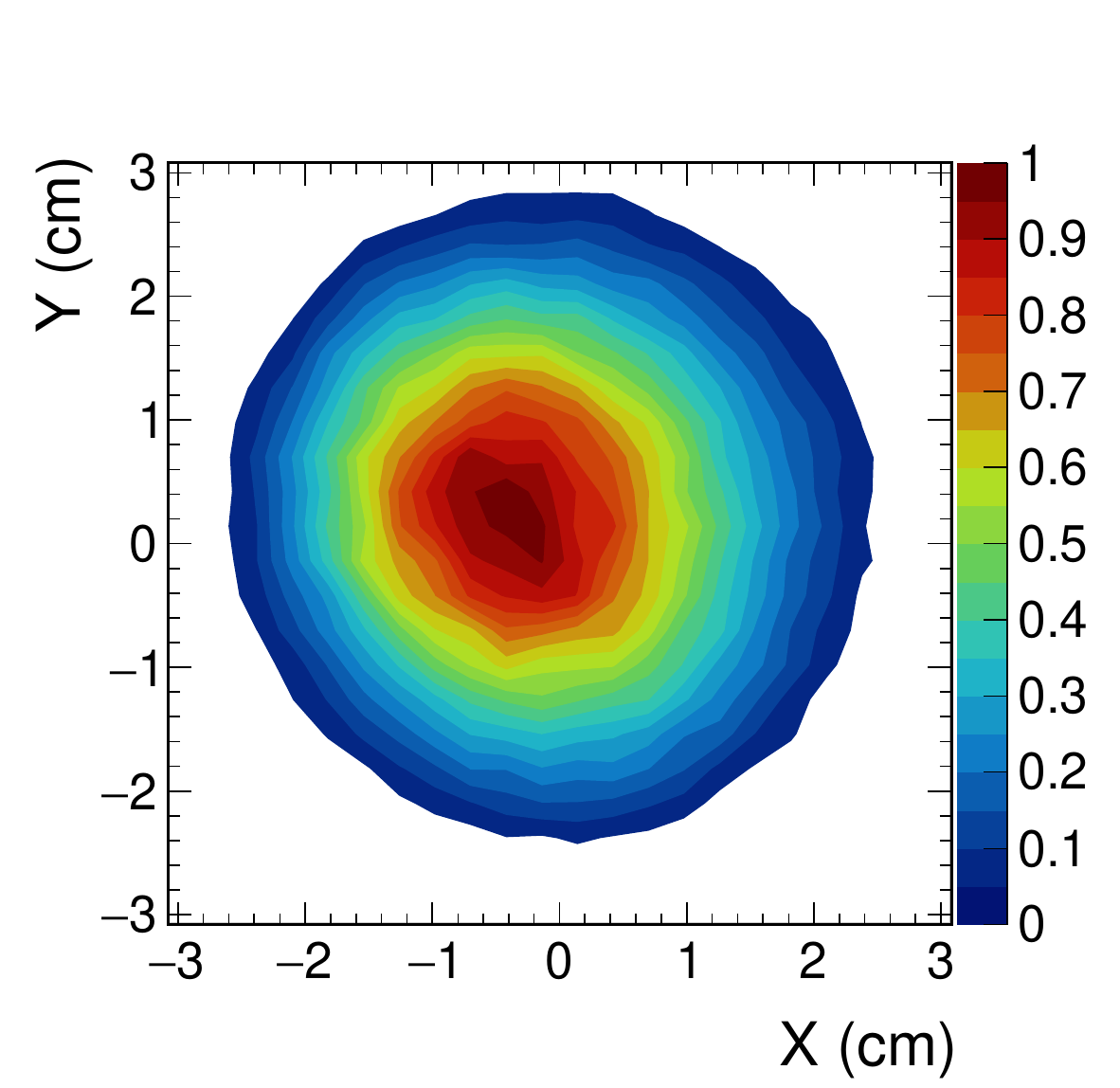}
\end{subfigure}
    \caption{Neutron beam profile measured with PPACMon at 19.95~m flight path in the energy range form 100 meV-100~keV (top) and from 100~keV to 100~MeV (bottom). The maximum of the distribution is scaled to 1.}
    \label{fig:5_profile_energy}       
\end{figure}

Figure~\ref{fig:5_profile_energy} then shows the neutron beam profile for two different energy ranges: from 10~meV to 100~keV, and from 100~keV up to 100~MeV. Both energy ranges show a beam profile (with approximately a Gaussian shape) with $\sim$ 3~cm FWHM. However, the actual beam shape and centre at these two energy ranges differs. This is due to the strong directionality of high-energy neutrons, mostly aligned with the direction of the proton beam, whereas lower-energy neutrons, due to moderation processes, have a more homogeneous distribution. 

\begin{figure}
\centering
\begin{subfigure}{.5\textwidth}
\centering
    \includegraphics[width=.9\linewidth]{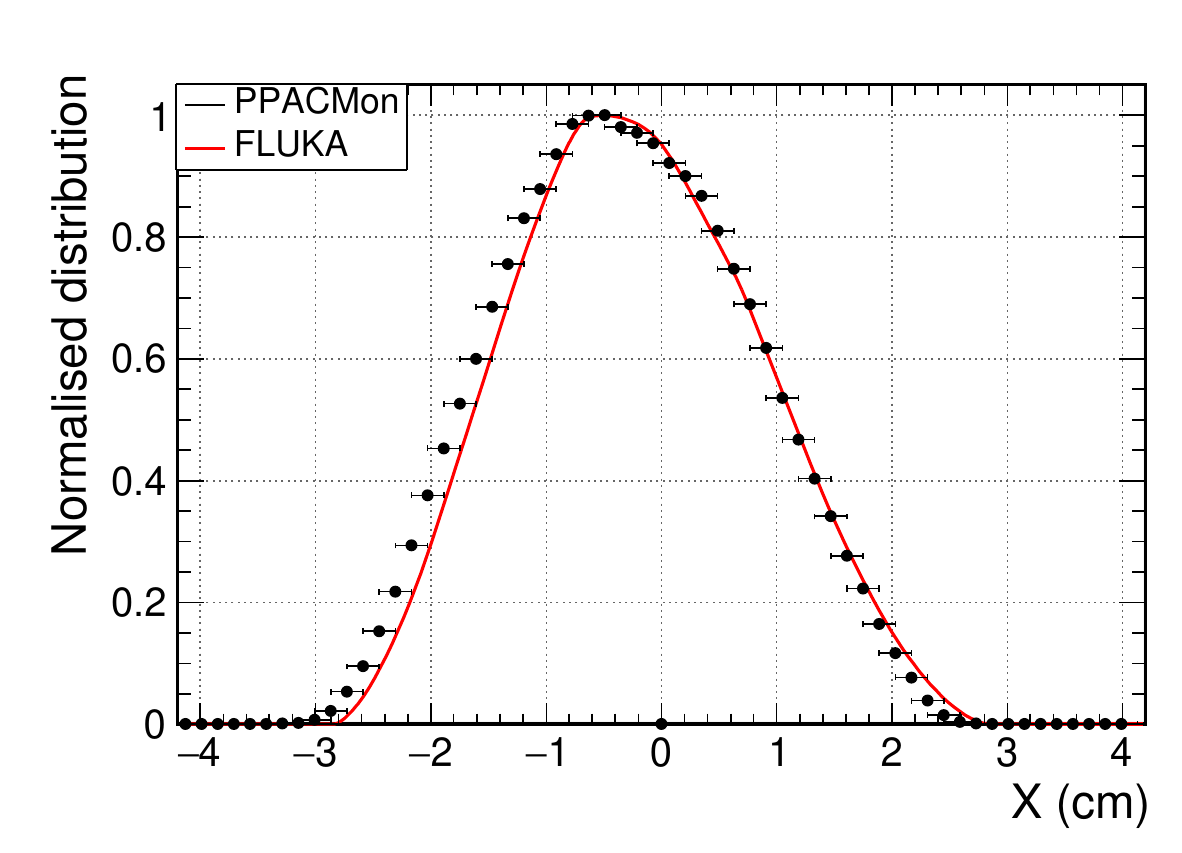}
\end{subfigure}
\begin{subfigure}{.5\textwidth}
\centering
    \includegraphics[width=.9\linewidth]{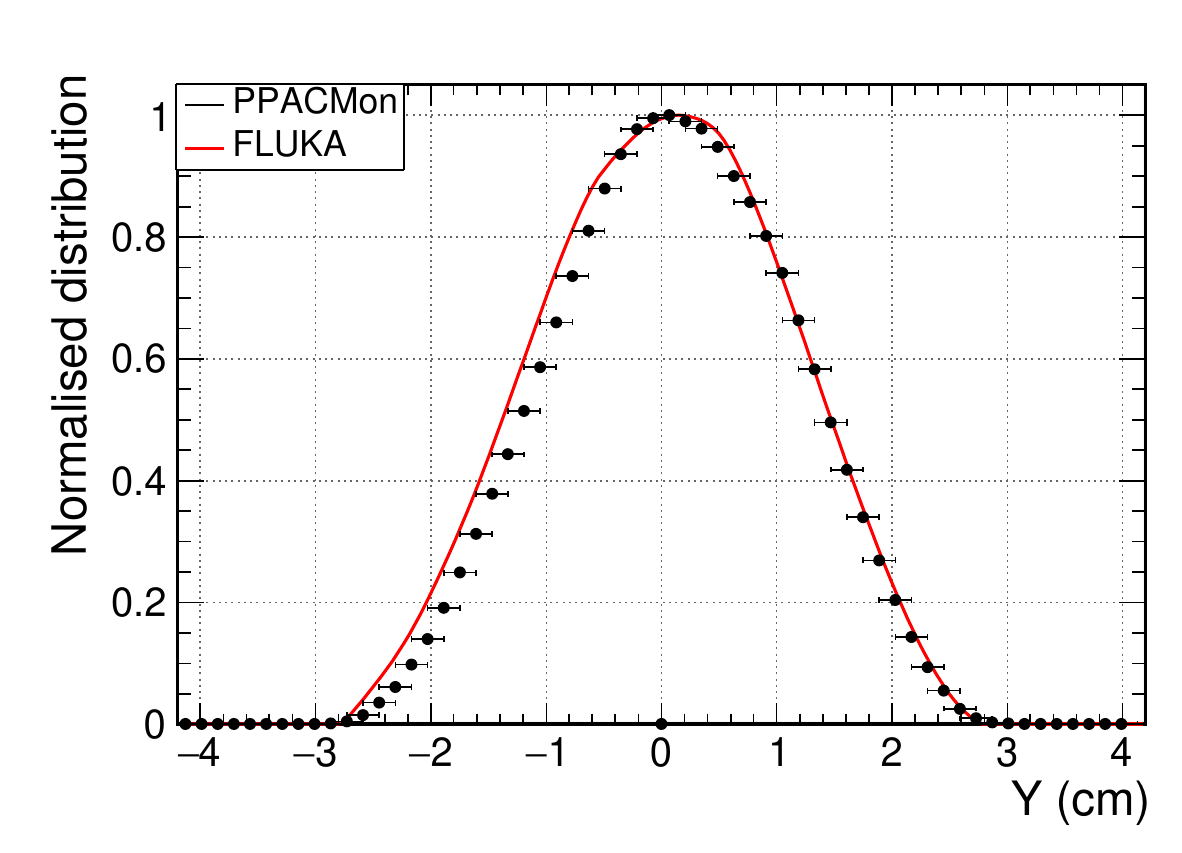}
\end{subfigure}
    \caption{Projection along the X-axis (top) and y-axis (bottom) at the peak value of the neutron beam profile measured with PPACMon at 19.95~m flight path (black dots) and from FLUKA simulations (red curve). The uncertainties in the experimental data are due to PPACMon's spatial precision of about 1.5~mm. The distribution are scaled to 1 at the maximum.}
    \label{fig:5_1d_flukaexp} 
\end{figure}

Figure~\ref{fig:5_1d_flukaexp} shows a comparison between PPACMon data and FLUKA simulations, of the projections along the $x$ and $y$ axes at the peak value of the neutron beam profile, integrated over the whole energy range. The asymmetries in the projection along the axis show that the beam does not have exactly a Gaussian shape. The simulations are in good agreement with the experimental data. 

\begin{figure}[ht!]
\centering
\begin{subfigure}{.5\textwidth}
\centering
    \includegraphics[width=0.95\linewidth]{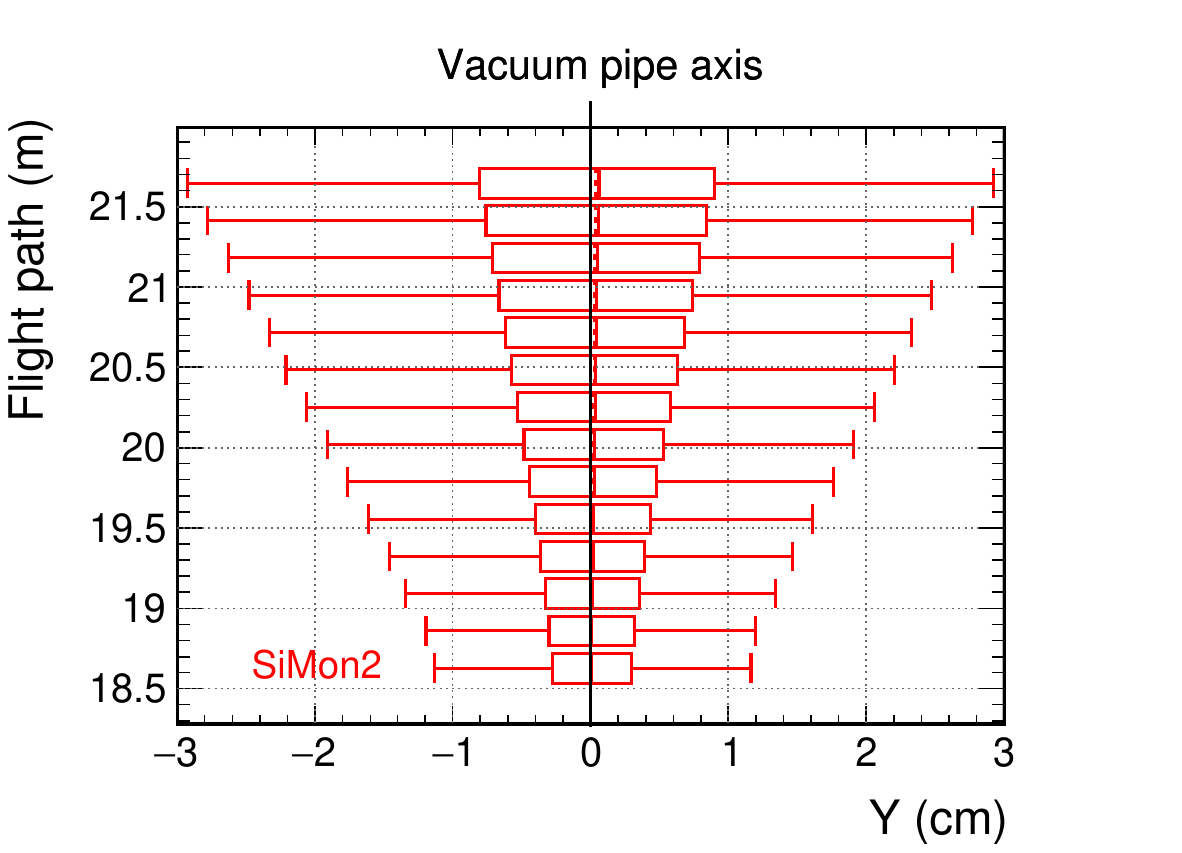}
\end{subfigure}
\begin{subfigure}{.5\textwidth}
\centering
    \includegraphics[width=0.95\linewidth]{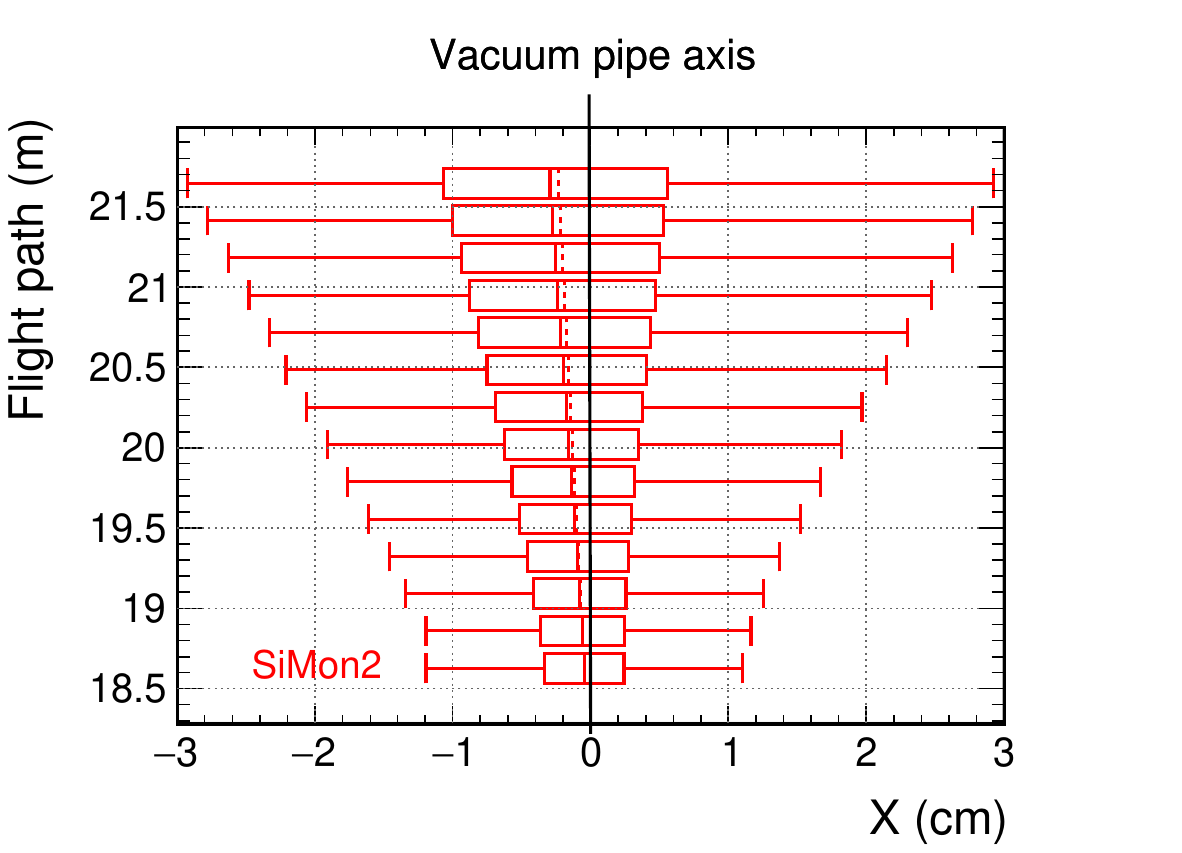}
\end{subfigure}
    \caption{Boxplot of the simulated beam profile, illustrating how the distribution changes along the Y-axis (top) and X-axis (bottom) with different flight paths inside the experimental hall. The central box represents the interquartile range ($\textrm{IQR}$) equivalent to $\pm0.675\times\sigma$ if the profile followed a Gaussian distribution, with the solid middle line indicating the median and the dashed line indicating the mean. The horizontal lines (whiskers) extend from $\textrm{IQR}/2$ to $1.5 \times \textrm{IQR}$, equivalent to $2.7\times\sigma$. The solid black line denotes the vacuum pipe axis.}
    \label{fig:5_centroid_profile}       
\end{figure}

Profiting from the simulations that reproduce well the observed spatial distribution, we have also simulated the beam profile inside EAR2 along the vertical axis, i.e. Z-axis. Figure~\ref{fig:5_centroid_profile} presents the profile distribution projected on the Y-axis (top) and X-axis (bottom) at different flight paths. The boxplot shows how the distribution spreads after entering in the experimental hall. The median of the distribution moves towards the negative x-axis with respect to the vacuum pipe axis as the beam moves up towards the ceiling. The displacement along the Y-axis at 19.95~m flight path is below 0.1~mm, while along the X-axis is $\sim$1.3~mm. Therefore, a precise alignment of the samples is required, especially for very small samples.

\begin{figure}
    \centering
\begin{subfigure}{.5\textwidth}
\centering
    \includegraphics[width=1.\linewidth]{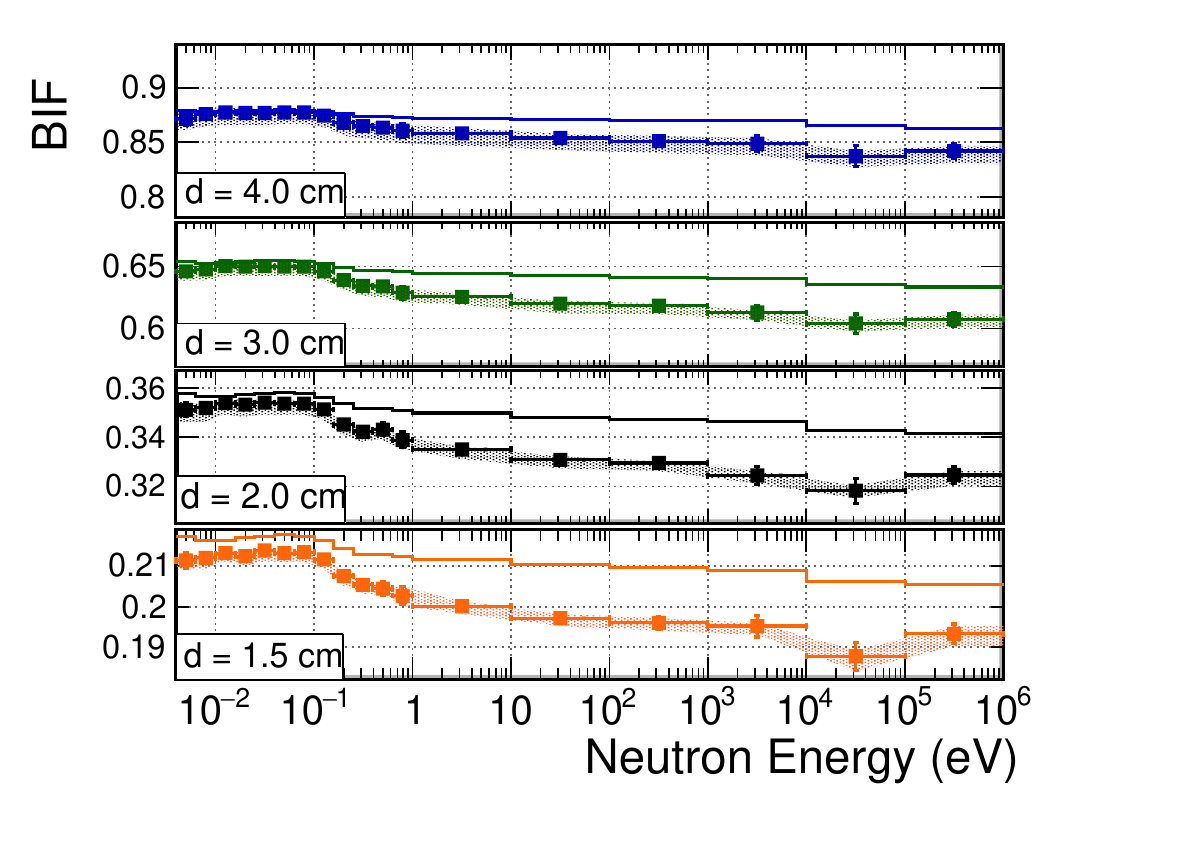}
\end{subfigure}
\begin{subfigure}{.5\textwidth}
\centering
    \includegraphics[width=1.\linewidth]{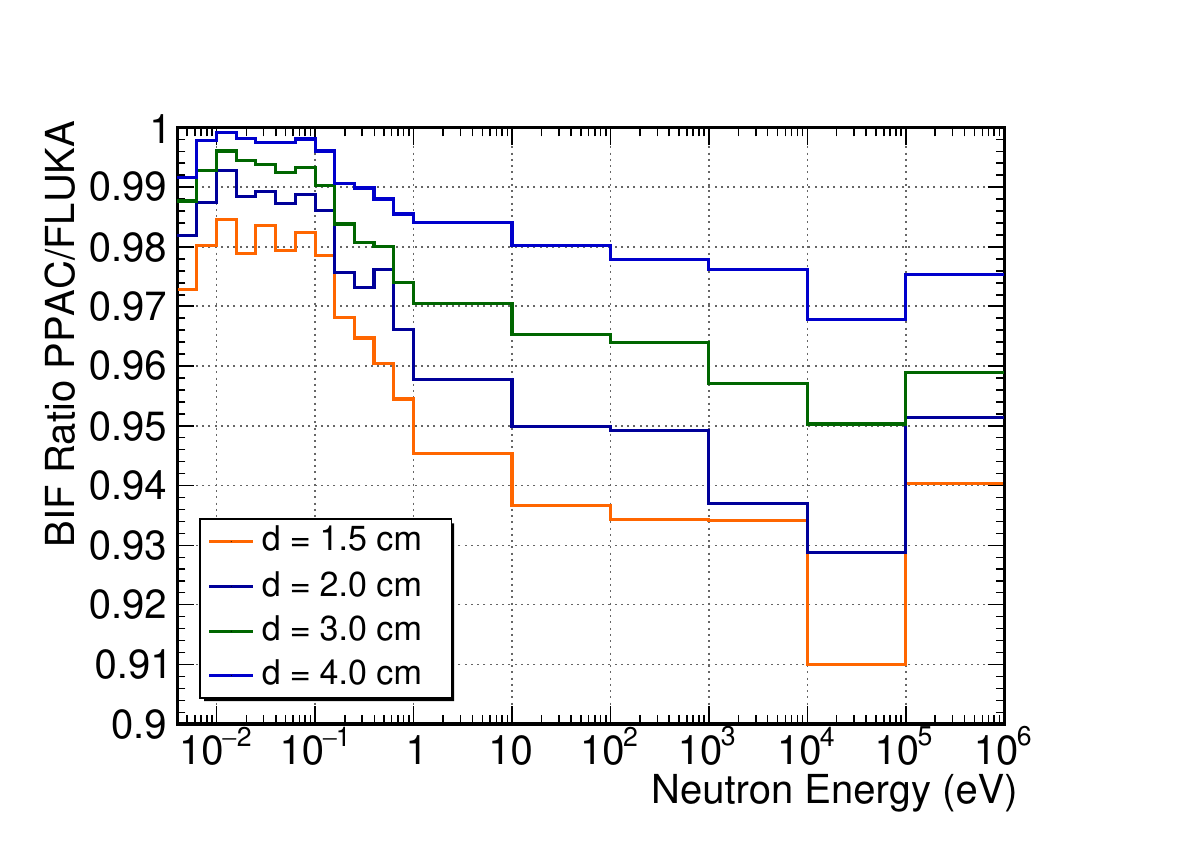}
\end{subfigure}
    \caption{Experimental and simulated BIF for different sample diameters at 19.95~m flight path (top) and ratio between PPACMon and FLUKA (bottom). In the top panel, the markers indicate the PPAC measurements while the solid lines are FLUKA simulations. The error bars represent the statistical uncertainty of the measurement, while the shaded area illustrates the effect of a misalignment of $\pm 1.4$~mm in the X and Y axes, which corresponds to PPACMon's spatial resolution.}
    \label{fig:5_BIF_profile}       
\end{figure}

Indeed, the accurate determination of the spatial profile is relevant, in particular, in the cross section measurement of samples which size is smaller than the neutron beam. In this case, only a fraction of the neutron beam is intercepted. This fraction is typically called the beam interception factor (BIF). The top panel of Figure~\ref{fig:5_BIF_profile} shows the experimental BIF (lines with markers) determined with PPACMon for samples of different diameters, aligned with the beam centre at thermal energy. The solid lines represent the BIF calculated with FLUKA for the same diameters. The larger BIF between 10 and 100~meV than for higher neutron energies is due to the choice of neutron energy range for the sample alignment. The bottom panel of Figure~\ref{fig:5_BIF_profile} presents the ratios between PPACMon and FLUKA, showing a sufficiently good agreement that can be exploited for the measurement planning. However, these differences must be considered if simulations are used to analyse measurements sensitive to the beam profile.

\section{Energy resolution}
\label{sec:RF}

A common characteristic of neutron TOF facilities is that neutrons of a given energy do not all exit the target-moderator assembly simultaneously, making the time-to-energy relation non-unique. As introduced in Sec.~\ref{sec:flux}, the relation between these two quantities is known as the energy resolution function (RF) of the facility and it can only be determined by means of MC simulations, which are then validated with experimental data.

To validate our FLUKA simulations of the RF, we have carried out measurements of (n,$\gamma$) reactions with $^{197}$Au and $^{56}$Fe samples, which show several observable resonance structures between a few eV and hundreds of keV with precisely known resonance parameters. The validation is  made by comparing the resonances measured experimentally with the ones obtained using the calculated RF with the R-Matrix analysis code SAMMY~\cite{webSammy}, which allows to include the effect of the RF to a calculated reaction yield.  

\begin{figure}
    \includegraphics[width=0.5\textwidth]{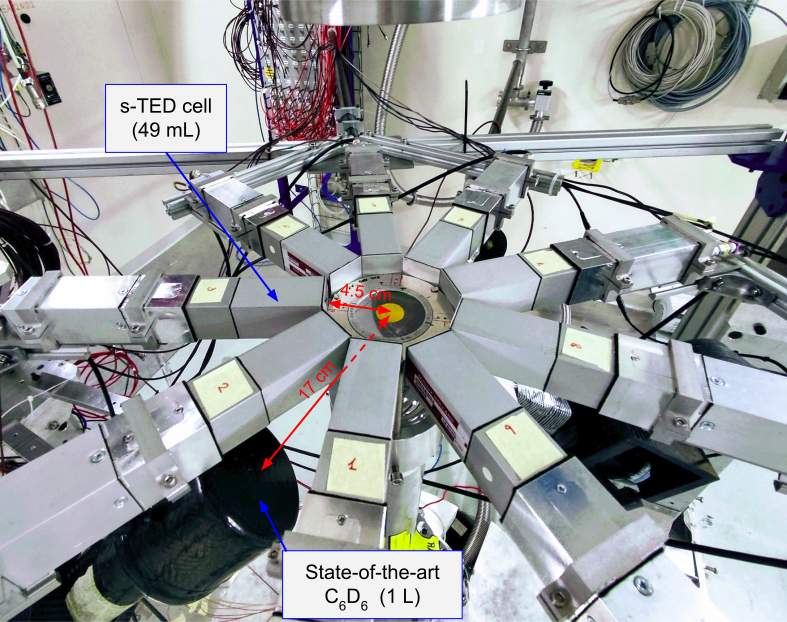}
    \caption{Experimental setup for (n,$\gamma$) measurements during the RF validation campaign.}
    \label{fig:6_rfsetup}       
\end{figure}

Figure~\ref{fig:6_rfsetup} shows the experimental setup, which consisted of two kinds of C$_{6}$D$_{6}$ scintillator detectors: two large-volume ones~\cite{Plag2003} and the sTED array consisting of nine small modules~\cite{Alcayne2024}. The former detectors contain 1 L of scintillating liquid in a carbon fiber case. In contrast, one sTED module contains only 49 ml of C$_{6}$D$_{6}$ in an aluminium case. Individual sTED modules are placed in an array of cylindrical symmetry around the sample, at the height of the reference position for capture measurements. The sTED configuration is a novel approach that profits from segmenting the liquid volume to better deal with high counting rates and a closer geometry that boosts the signal to background ratio~\cite{Lerendegui2023arxiv}.

\begin{figure}
\centering
\begin{subfigure}{.5\textwidth}
\centering
    \includegraphics[width=1.\linewidth]{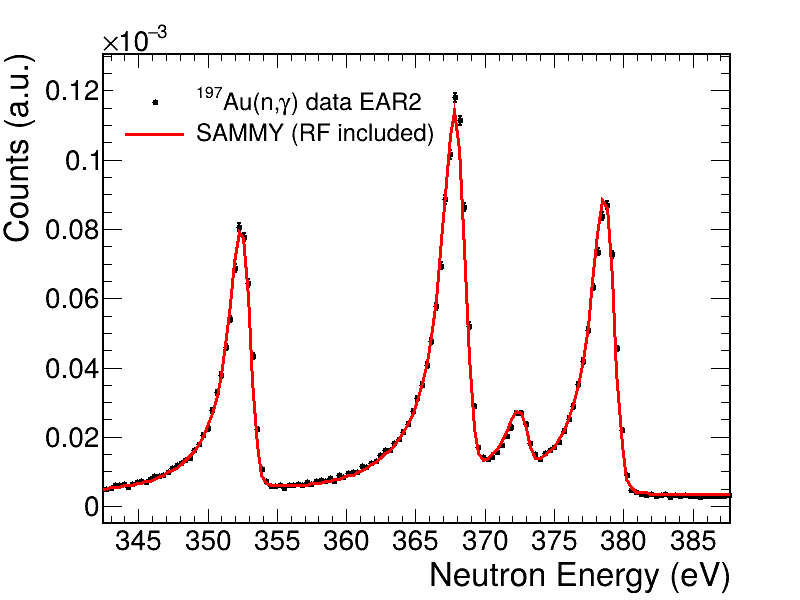}
\end{subfigure}
\begin{subfigure}{.5\textwidth}
\centering
    \includegraphics[width=1.\linewidth]{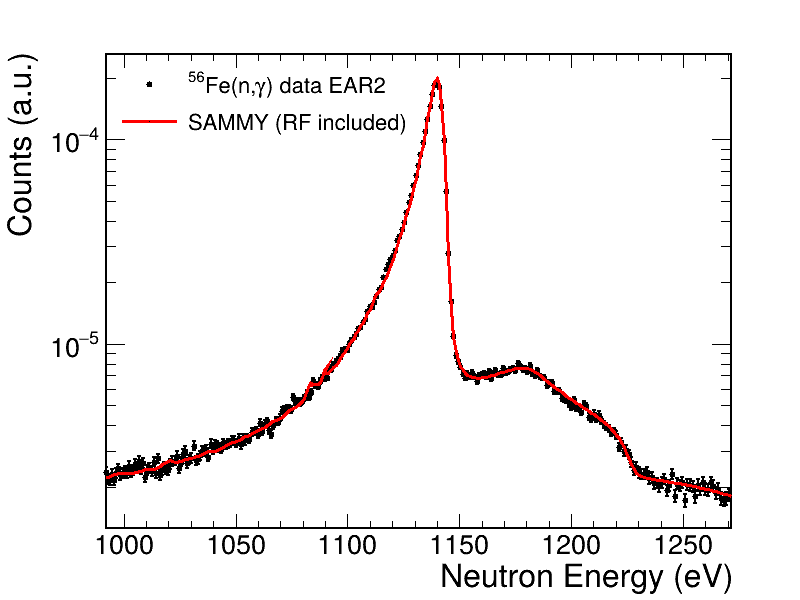}
\end{subfigure}
    \caption{Experimental resonances of $^{197}$Au(n,$\gamma$) (top) and $^{56}$Fe(n,$\gamma$) (bottom) compared to the predictions of SAMMY including the FLUKA RF~\cite{Lerendegui2023arxiv}.}
    \label{fig:6_RFvalidation}       
\end{figure}

Figure~\ref{fig:6_RFvalidation} shows a comparison between measured resonances of $^{197}$Au(n,$\gamma$) and $^{56}$Fe(n,$\gamma$) and the calculation with SAMMY~\cite{webSammy} using ENDF/B-VIII.0 resonance parameters and our calculated RF extracted from FLUKA simulations. The agreement between experimental data and the calculated yield, without any parameter fitted, is remarkable. It demonstrates the ability of the calculated RF to reproduce both the broadening and the energy shift of the resonances. We checked the high quality of the reproduction up to several tens of keV.  

\begin{figure}
\centering
\begin{subfigure}{.5\textwidth}
\centering
    \includegraphics[width=1\linewidth]{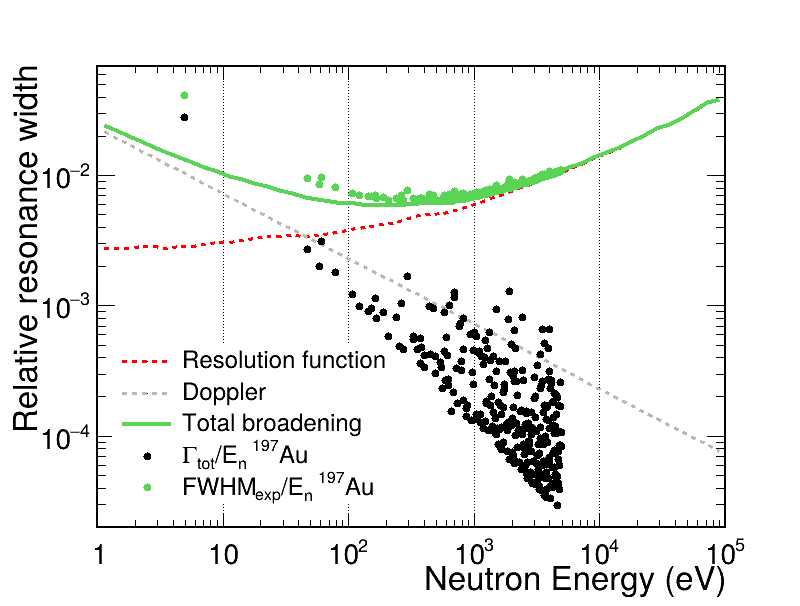}
\end{subfigure}
\begin{subfigure}{.5\textwidth}
\centering
    \includegraphics[width=1\linewidth]{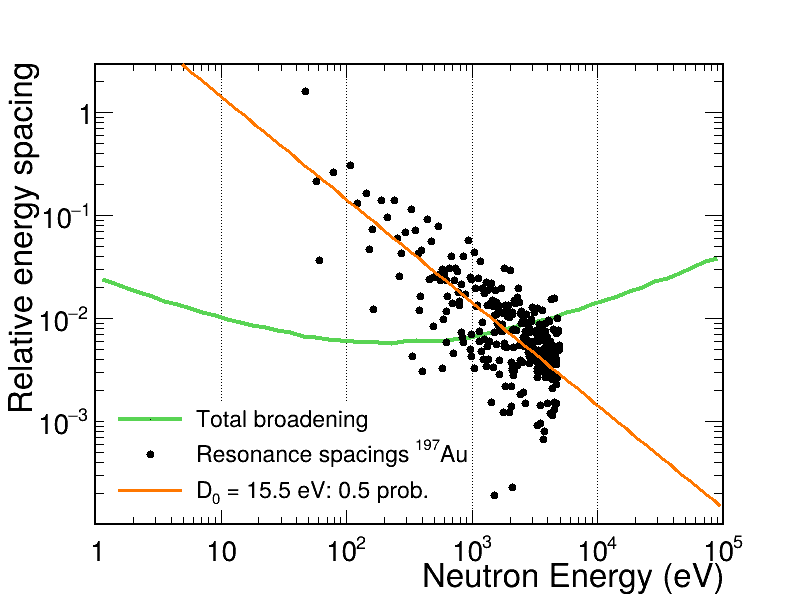}
\end{subfigure}
\caption{Top: Relative resonance widths of $^{197}$Au resonances ($\Gamma_{tot}$/$E_{n}$) as a function of the neutron energy. The contribution of individual broadening effects (Doppler effect, RF) to the experimentally observed relative width FWHM$_{exp}$/$E_\textrm{n}$ is shown. Bottom: Relative energy spacing between neighbouring $^{197}$Au resonances compared to the total relative observed resonance width. The orange line corresponds to the median of the s-wave spacing distribution (see text for further details).}
    \label{fig:6_RFimpact}  
\end{figure}

To illustrate the impact of the RF in the broadening of the resonances, in the top panel of Figure~\ref{fig:6_RFimpact} the relative intrinsic widths of the $^{197}$Au(n,$\gamma$) resonances, $\Gamma_\textrm{tot}/E_\textrm{n}=(\Gamma_\textrm{n}+\Gamma_{\gamma})/E_\textrm{n}$, in the JEFF-3.3 library (black dots) are compared to the experimental broadening contributions corresponding to the relative width at half maximum (FWHM/$E_\textrm{n}$). In this figure, the overall resonance broadening is displayed as a thick (green) solid line, while the individual components (Doppler effect and RF) are displayed as dashed lines. The Doppler broadening is the main contribution at neutron energies $E_\textrm{n}<50$~eV, while the resolution broadening becomes the most important at higher energies. The total width for the observed $^{197}$Au(n,$\gamma$) resonances (green dots) is fully dominated by the experimental broadening for $E_\textrm{n}\geq100$~eV.  

The high resolution of the n\_TOF-EAR2 neutron beam makes possible to analyse individual resonances in a wide energy range and determine the cross section from the obtained resonance parameters. At higher energies, where the resonances are not well resolved, another approach must be used which is more sensitive to systematic uncertainties in the determination of the background. As visible from the top panel of Figure~\ref{fig:6_RFimpact}, the RF often represents the main contribution to the observed resonance width and thus the main limitation for determining the cross section from individual resonance parameters, since it limits the ability to disentangle neighbouring resonances.

To illustrate the resolving power at EAR2, the bottom panel of Figure~\ref{fig:6_RFimpact} shows the energy spacing between neighbouring $^{197}$Au(n,$\gamma$) resonances from the JEFF-3.3 library divided by the energy $E_\textrm{n}$ of the first of these resonances (black dots). The green curve in this figure represents the total experimental broadening, which is representative of the total resonance width for $^{197}$Au. Known resonances below the green line are separated by an energy smaller than the observed resonance width, and thus are overlapped with their neighbour resonance. To indicate the impact of this overlap we added the orange straight line to the figure. For each neutron energy, the line corresponds to $0.95\times D_0$, for which 50\% of all the spacings lie below the line according to the expected Wigner distribution of nearest s-wave resonance spacings (D$_0$ = 15.5~eV~\cite{RIPL}). The energy of the intersection between the orange lines and the green curve, around 1.5~keV, then indicates the energy at which about a half of the resonances will remain unresolved from their neighbouring resonances.  As indicated by the data of the experimental commissioning, this neutron energy really represents the upper limit for the analysis of resolved $^{197}$Au(n,$\gamma$) resonances in n\_TOF-EAR2. 

\begin{figure}
\centering
\begin{subfigure}{.5\textwidth}
\centering
    \includegraphics[width=1.\linewidth]{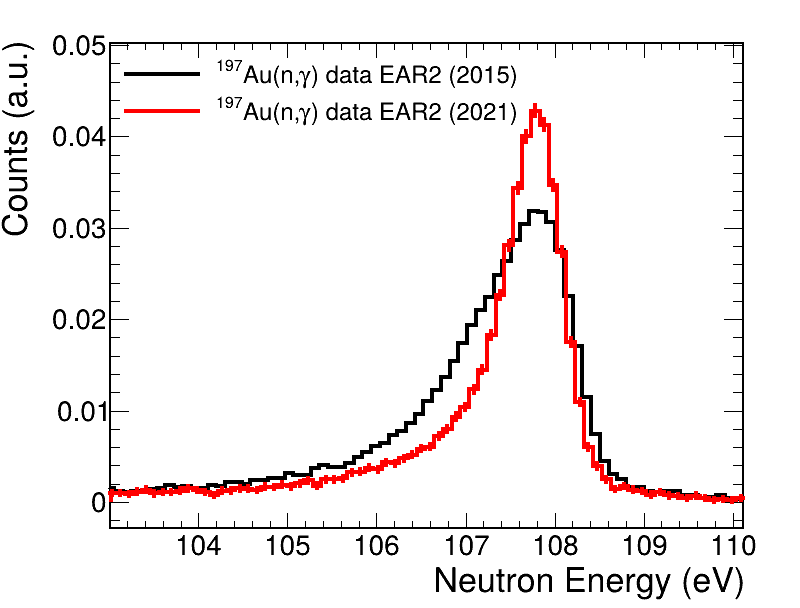}
\end{subfigure}
\begin{subfigure}{.5\textwidth}
\centering
    \includegraphics[width=1.\linewidth]{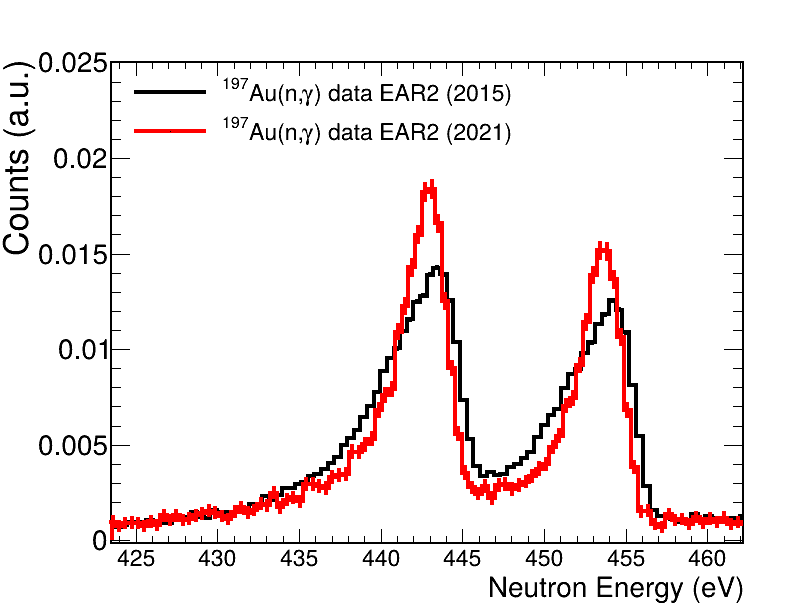}
\end{subfigure}
    \caption{Capture resonances measured at different neutron energies with the old(2015)and new(2021) spallation target. Both data sets have been normalized to the resonance area. The improvement in the resolution is visible in the narrower shape of the resonance.}
    \label{fig:6_RFphase3n4}       

\end{figure}

The energy resolution at EAR2 has significantly improved after the installation of the new spallation target and it is no longer affected by the precise alignment of the sample~\cite{rfcapture}. Figure~\ref{fig:6_RFphase3n4} illustrates the improved resolution by comparing $^{197}$Au(n,$\gamma$) resonances measured at EAR2 with the previous spallation target (2015) and the new target (2021). Resonances at higher energies in EAR2 can thus be resolved with the new spallation target. An improved energy resolution is also a key aspect for both increasing the signal-to-background ratio and obtaining more accurate resonance parameters. Table~\ref{tab:rfvalues} summarizes the relative energy resolution (FWHM) as a function of the neutron energy and compares it with the results of the previous spallation target.
\begin{table}
\centering
\caption{\textcolor{black}{Energy resolution $\Delta E_\textrm{n}/E_\textrm{n}$ at FWHM as a function of neutron energy for the new spallation target compared to the previous target }.}
\label{tab:rfvalues}
 \begin{tabular}{lcc}
    \hline
    Neutron energy   & New target                  & Previous target  \\
                     & $\Delta E_\textrm{n}/E_\textrm{n}$            & $\Delta E_\textrm{n}/E_\textrm{n}$ \\\hline
    10 meV           &   $7.1   \times 10^-3 $     &  $1.3   \times 10^-2 $    \\   
    100 meV          &   $5.4   \times 10^-3 $     &  $1.4   \times 10^-2 $    \\   
    1 eV             &   $2.8   \times 10^-3 $     &  $4.8   \times 10^-3 $    \\   
    10 eV            &   $3.1   \times 10^-3 $     &  $5.1   \times 10^-3 $    \\   
    100 eV           &   $3.8   \times 10^-3 $     &  $7.1   \times 10^-3 $    \\   
    1 keV            &   $6.0   \times 10^-3 $     &  $1.3   \times 10^-2 $    \\   
    10 keV           &   $1.4   \times 10^-2 $     &  $2.3   \times 10^-2 $    \\   
    100 keV          &   $4.2   \times 10^-2 $     &  $4.5   \times 10^-2 $   \\     
    1 MeV            &   $5.8   \times 10^-2 $     &  $5.7   \times 10^-2 $   \\     
    \hline
\end{tabular}
\end{table}

The newly upgraded resolution of n\_TOF-EAR2 is compared to other TOF facilities worldwide featuring white neutron beams and similar flight paths in Table~\ref{tab:rfvsothers}. This table summarises for all the listed facilities the relative energy resolution (FWHM) at two neutron energies, 1~eV and 1~MeV. The results show how after the upgrade, n\_TOF-EAR2 presents one of the bests resolutions at low neutron energies (1~eV) among the facilities with similar flight paths. The values indicated for DANCE will be significantly improved after the recent upgrade of the spallation target to Mark-IV~\cite{DANCE}. As for ANNRI and CNSN back-n, the values in Table~\ref{tab:rfvsothers} do not consider the broadening due to the double bunch structure. 

\begin{table}
\centering
\caption{\textcolor{black}{Energy resolution $\Delta E_\textrm{n}/E_\textrm{n}$ at FWHM for $E_\textrm{n}$=1~eV and $E_\textrm{n}$ = 1~MeV for EAR2 compared to existing TOF facilities with similar flight path (L), as well as with n\_TOF-EAR1.}}
\label{tab:rfvsothers}
 \begin{tabular}{lccc}
    \hline
               &        & \multicolumn{2}{c}{$\Delta E_\textrm{n}/E_\textrm{n}$} \\
      Facility &  L (m) & $E_\textrm{n}$=1~eV &$E_\textrm{n}$=1~MeV       \\
          \hline
      n\_TOF-EAR2 &   20    &   2.8$\times 10^-3$    &     5.8$\times 10^-2$    \\
       GELINA~\cite{GELINA}      &  32    &   1.3$\times 10^-3$    &     3.5$\times 10^-3$    \\
      DANCE~\cite{DANCE}        &   20    &  3.5$\times 10^-3$      &    2.5$\times 10^-2$    \\
     ANNRI~\cite{ANNRI}         &   21.5  &  3.5$\times 10^-3$     &     9.0$\times 10^-2$               \\
    CNSN BACK-N~\cite{BACKN}   &  56      &  4.0$\times 10^-3$      &     2.2$\times 10^-2$               \\
     n\_TOF-EAR1~\cite{Guerrero2013}  &   185     &  3.2$\times 10^-4$       &  5.4$\times 10^-3$                 \\
    \hline
\end{tabular}
\end{table}

\section{Conclusions}
\label{sec:conclusions}

The n\_TOF-EAR2 neutron beamline required a full characterisation after the installation of the new spallation target in 2021 during the CERN $2^{\textrm{nd}}$ Long Shutdown, necessary for the analysis of the measurements in the facility. In this paper, we presented the methodology and results of the commissioning phase during which we determined the neutron flux in a very broad energy range and the spatial beam profile for the small collimator, together with the energy resolution function of this beamline. 

The facility exhibits a high-quality neutron beam, characterised by broad energy spectrum spanning from below 10~meV up to 200~MeV and high instantaneous intensity per energy decade between $0.6-4.3 \times 10^{6}$~neutron/pulse. Compared with the previous target, the neutron flux is higher by about 45\% in the epithermal region (1~eV to 100~keV), and by about 20\% in the thermal region (below 0.5~eV). In the evaporation peak (100~keV to 10~MeV) the flux increase is of approximately 40\%. The statistical uncertainties in the measurements presented here in 100 bins per decade remain below the systematic ones for the entire energy range with the exception of the highest energies ($E_\textrm{n}>1$~MeV). The systematic uncertainties remain below 3\% up to 1~MeV and below 5\% at higher energies.

The Monte Carlo simulations of the neutron beam carried out with FLUKA showed a very good agreement in shape with the evaluated flux in the epithermal region, which was within 3\% of the evaluation's uncertainties. In the thermal region and for $E_\textrm{n}>100$~keV the agreement was within about 10\%. This deviation can be due to imperfections in the collimation (and small imperfections in the geometry model  implemented in the simulations). 

The spatial beam profile measured with PPACMon at 19.95~m flight path shows an FWHM of about 3~cm. The predictions of the beam profile of the FLUKA simulations at this position showed a remarkable agreement with the measurements. In addition, we investigated the variation of the beam interception factor with energy for a wide range of sample dimensions. We found that this factor is at most within ~5\% between different energy decades. Moreover, the agreement with FLUKA predictions was found to be between 2 to 9\%, which could be due to the collimation system and/or to the relative alignment between PPACMon and the FLUKA coordinates. 

The FLUKA simulations were also used to calculate the energy resolution function of the facility. The resolution function of the EAR2 beamline is significantly narrower compared to the previous spallation target. The use of this calculated resolution function in the SAMMY R-matrix code perfectly reproduces the resonance shapes observed in the $^{197}$Au(n,$\gamma$) and $^{56}$Fe(n,$\gamma$) experimental data for neutron energies up to tens of keV. This improvement will allow to determine resonance parameters of observed resonances with a higher precision and over a wider energy range than with the previous spallation target. If we consider the reference neutron capture reaction on $^{197}$Au, the resolution function available at EAR2 should now allow resolving individual resonances up to a neutron energy of about 1.5~keV. After the recent upgrade, n\_TOF-EAR2 presents one of the best resolutions at low neutron energies (e.g. 1~eV) among the TOF facilities with comparable flight paths.   

The new capabilities of the n\_TOF-EAR2 facility presented here, along with the advanced detector systems currently under development~\cite{Alcayne2024, Balibrea2024}, provide an excellent foundation for conducting new and more challenging physics measurements. Notable examples include the first-ever cross section measurements on mg samples of radioactive isotopes such as $^{94}$Nb(n,$\gamma$)~\cite{Balibrea2023} and $^{79}$Se(n,$\gamma$)~\cite{Lerendegui-Marco2023det}, as well as new measurements of the $^{146}$Nd(n,$\gamma$)~\cite{Lerendegui-Marco2023} and $^{243}$Am(n,f) cross sections~\cite{Patronis2020}, among numerous other ongoing measurements included in the experimental programme of the facility.
\section*{Acknowledgments}
This work was financially supported by Grant PID2021-098117-B-C21 funded by MCIN/AEI/10.13039/501100011033 and by “ERDF A way of making Europe”. Additionally, the authors thank the support of postdoctoral grants FJC2020-044688-I and  CIAPOS/2022/020 funded, respectively, by MCIN AEI/10.13039/501100011033 and Generalitat Valenciana. Support from the funding agencies of all other participating institutes is gratefully acknowledged. 
The authors thank the n\_TOF Collaboration members for their support during the commissioning campaign.
\bibliographystyle{elsarticle-num-names}
\bibliography{references.bib}

\end{document}